\documentclass{imsart}

\RequirePackage[OT1]{fontenc}
\RequirePackage{amsthm,amsmath}
\RequirePackage[numbers]{natbib}
\RequirePackage{natbib}
\RequirePackage[colorlinks,citecolor=blue,urlcolor=blue]{hyperref}

\usepackage{amssymb}
\usepackage{graphicx}
\usepackage{amsmath}
\usepackage{rotating}
\usepackage{lscape}
\usepackage{rotfloat}
\usepackage{subfig}
\usepackage{dsfont,srcltx}
\usepackage{amsfonts}
\usepackage[bottom]{footmisc}
\usepackage{amsfonts}
\usepackage{dsfont,srcltx}
\usepackage[english]{babel}
\usepackage{multirow}
\usepackage{bbm}
\usepackage{amsmath,amssymb,amsthm,mathtools}
\usepackage{amsfonts}
\usepackage{graphicx,dsfont,srcltx}
\usepackage{bbm}
\usepackage{natbib}
\usepackage{hyperref}
\usepackage{enumitem}
\usepackage{booktabs}
\usepackage{placeins}

\hypersetup{
  colorlinks=true,
  linkcolor=blue,
  citecolor=blue,
  urlcolor=blue
}
\theoremstyle{plain}% default
\newtheorem{theorem}{Theorem}[section]

\newtheorem*{remark}{Remark}
\theoremstyle{definition}

\newtheorem{example}{Example}[section]
\newtheorem{assumption}{Condition}
\newtheorem*{assumptionbis}{Condition C4'}
\allowdisplaybreaks[4]

\newcommand{\F}{\mathcal F}
\newcommand{\X}{\mathcal X}

\newcommand{\E}{\mathbb E}

\newcommand{\R}{\mathbb R}
\newcommand{\1}{\mathbbm 1}

\newcommand{\weak}{\Rightarrow}

\newcommand{\var}{\mathbb{V}}

\begin{document}
\begin{frontmatter}
\title{A General Theory of Nonparametric Response-Adaptive Randomization}
\runtitle{Nonparametric Response-Adaptive Randomization}
\begin{aug}
\author[A]{\fnms{Alessandro}~\snm{Baldi Antognini}
\ead[label=e1]{a.baldi@unibo.it}}
\author[A]{\fnms{Rosamarie}~\snm{Frieri}
\ead[label=e2]{rosamarie.frieri2@unibo.it}}
\author[A]{\fnms{Maroussa}~\snm{Zagoraiou}
\ead[label=e3]{maroussa.zagoraiou@unibo.it}}
\runauthor{A. Baldi Antognini, R. Frieri and M. Zagoraiou}
\address[A]{Department of Statistical Sciences,
University of Bologna, Bologna, Italy.\\
\printead[presep={E-mail:\ }]{e1,e2,e3}}
\end{aug}

\begin{abstract}
Response-Adaptive Randomization (RAR) assigns incoming patients to treatments with probabilities that depend on previously observed responses. While most existing RAR procedures rely on finite-dimensional summaries of treatment performance (such as rates, means, or model parameters), we develop a nonparametric (NP) distribution-based framework in which the treatment outcome distributions are learned sequentially and adaptation is driven by their functionals. The framework naturally accommodates multiple treatments and is particularly suited to endpoints for which average effects may conceal relevant differences across the outcome distribution. We introduce an ethical NP-RAR rule driven by a pre-specified distributional score and a nonparametric extension of the multi-arm efficient randomized adaptive design (NP-ERADE) able to approach generic distribution-dependent targets selected according to ethical or inferential optimality criteria. We establish strong consistency and joint functional central limit theorems for the empirical treatment distributions, together with asymptotic normality of functional scores, estimated targets and allocation proportions. We further show that the NP-ERADE allocation proportions attain the semiparametric efficiency bound for the target. We develop post-trial inference for score contrasts and distributional features, including crossing locations, and investigate finite-sample properties, robustness and inferential performance via simulations.
\end{abstract}



\end{frontmatter}

\section{Introduction}
Response-adaptive randomization (RAR) is a class of sequential experimental designs in which the probability of assigning a new statistical unit to a treatment is updated as outcome information accrues. RAR procedures have been developed mainly for two distinct purposes: to address ethical considerations in clinical trials by progressively favoring treatments that appear superior while retaining randomization and preserving inferential validity, or to solve local optimal-design problems by targeting allocation proportions depending on unknown model parameters. The classical literature includes play-the-winner rules, generalized urn designs, doubly adaptive biased coin designs, efficient randomized-adaptive designs (ERADE) and Bayesian procedures (see, among others, \cite{giova,Hu03,Hu06,Hu04,Hu09,WeiDur78,Zel69,Zha06}). More recently, RAR procedures have also been developed in a bandit perspective, with particular emphasis on non-myopic, forward-looking rules that trade off learning about treatment effects and the ethical requirements for the enrolled patients (see \cite{villar2015,VWB15,Villar20}). Recent work emphasizes both the renewed methodological interest in RAR and the practical concerns that have limited its routine use, including power, time trends, imbalance, delayed responses, type-I error control and regulatory scrutiny \citep{baas2024exact,robertson2023myths,TangBaasKaddajPinRobertsonVillar2025}.

A common feature of most RAR methodologies is that the accumulating responses affect the design through finite-dimensional treatment summaries (typically, the estimated treatment effects). However, it can be restrictive when the relevant comparison between treatments is distributional rather than mean-based. Many outcomes are continuous, ordinal, semicontinuous, or composite, and treatment effects may differ across severity levels, tails, or clinically critical regions. A therapy may improve the lower tail of a quality-of-life score while leaving the mean nearly unchanged, reduce the probability of severe deterioration while increasing variability, or benefit moderate patients while harming frail ones. In such cases, RAR designs driven by means, response rates, or parametric working models may obscure the very features that are relevant to the treatment comparison. Examples include quality-of-life, where lower tails represent severe disability; biomarker responses, where improvement near a therapeutic threshold matters more than improvement elsewhere; pain or symptom scales, where effects may differ across severity levels. In these settings, the treatment distributions should be regarded as the primary statistical objects, while the functional score translates distributional evidence into a treatment ranking and, in turn, into an adaptive allocation probability.

Moreover, the vast majority of the RAR literature is developed under the implicit or explicit assumption that treatment responses belong to a common parametric family and that treatment superiority can therefore be summarized through a finite-dimensional parameter, typically the mean (see, among others, \cite{baldi2015adaptive,Hu03,Hu06,Hu04,Hu09}). This perspective is naturally supported in the one-parameter exponential family, where the ordering induced by the canonical parameter is equivalent to the ordering of the means, so that ranking treatments according to their effects is coherent with the underlying model structure. However, this equivalence does not hold in general. Even within standard parametric models, different means do not necessarily imply a distributional ordering: under heteroscedastic normal responses, for instance, two treatments may have different means while their cumulative distribution functions still cross, so that neither treatment dominates the other in the stochastic-order sense. More importantly, in many applications the response distributions under different treatments may not even belong to the same parametric family, or any common distributional specification may simply be unsupported by the available scientific knowledge. In this case, even the use of a common mean-based treatment ranking lacks the distributional justification provided by a shared parametric family. Moreover, even when treatment performance is defined through the same clinically relevant functional, a misspecified parametric model may asymptotically reverse the treatment ranking.

The aim of this paper is to develop a nonparametric theory linking adaptive allocation, inference on treatment distributions and semiparametric efficiency. Without specifying parametric models for the treatment responses, we define treatment comparisons and allocation targets through distributional functionals. This formulation accommodates multiple treatments and heterogeneous response distributions, with objectives involving quantiles, tail behavior, threshold attainment or integrated distributional criteria. The joint empirical-process theory provides a common basis for both design and post-trial inference. The paper makes several contributions. First, in Section \ref{Sec3} we introduce an ethical nonparametric distribution-based RAR procedure (NP-RAR), driven by sequentially estimated treatment-specific cumulative distribution functions (CDFs). In Section \ref{Sec4} we establish its large-sample theory, proving strong consistency of the empirical CDFs, scores, assignment probabilities and allocation proportions. We derive a joint functional central limit theorem for the empirical CDFs and the asymptotic distributions of the scores and allocation probabilities. We also establish asymptotic normality of the allocation proportions, with an explicit variance decomposition separating intrinsic randomization variability from the additional variability generated by sequential learning. Second, in Section \ref{Sec5} we derive $A$-, $D$- and $E$-optimal distribution-based targets for estimation of the treatment scores and develop functional $A$- and $E$-optimality criteria for joint estimation of the complete treatment distributions. Third, we propose a nonparametric extension of the multi-arm efficient randomized adaptive designs (NP-ERADE) aimed at converging to distribution-based targets and we establish its large-sample theory (Section \ref{Sec6}). A central result shows that, under some regularity conditions, the allocation proportions of the proposed NP-ERADE attain the semiparametric efficiency bound for estimating the functional target. Section \ref{Sec7} discusses post-trial inference on the pre-specified scores and on the complete CDFs, including distributional contrasts and crossing locations. Finally, Section \ref{Sec8} investigates the finite-sample behavior, robustness and inferential performance of the proposed procedures through simulations and Section \ref{Sec9} concludes with a discussion.

The key conceptual distinction from earlier nonparametric adaptive randomization literature is the use of the empirical treatment distributions as the objects driving both the adaptation process and post-trial inference. While previous proposals adopt nonparametric approaches for clinical trials in the presence of covariates, with the aim of improving the similarity of the covariate distribution across arms \cite{MaHu12} or for constructing covariate-specific allocation targets \cite{aletti2018nonparametric}, Liu et al. \cite{LiuHuZhang2018} proposed a nonparametric RAR procedure for continuous outcomes with two treatments based on the Mann-Whitney U-statistic. Their approach provides an important distribution-free alternative to parametric RAR, but relies on a specific pairwise comparison criterion. Closer to the present perspective, Kock et al. \cite{KockPreinerstorferVeliyev2022} studied sequential allocations based on functionals of the outcome distributions in a multi-armed bandit framework. Their objective is cumulative regret minimization, so that exploration is instrumental to identifying and increasingly exploiting the treatment maximizing the chosen functional. By contrast, our framework is fundamentally different and remains within the classical RAR paradigm: treatment assignment remains randomized and all arms receive a non-vanishing asymptotic fraction of observations, thereby preserving simultaneous learning of all treatments and allowing inference for all distributions and their scores (accordingly, our proposal is not intended to be regret-optimal in the bandit sense).

\section{Setup and notation}\label{Sec2}
Consider a sequential experiment with $K\ge2$ treatments, indexed by $k=1,\ldots,K$. Let $\boldsymbol{\delta}_i=(\delta_{i1},\ldots,\delta_{iK})^\top$ be the treatment allocation of the $i$th statistical unit, with $\delta_{ik}=1$ if this subject is assigned to treatment $k$ and $0$ otherwise, where $\sum_{k=1}^K\delta_{ik}=1$ for every $i$. Moreover, let $Y_{ik}$ denote its potential outcome under treatment $k$, then the observed outcome is $Y_i=\sum_{k=1}^K\delta_{ik}Y_{ik}$, with larger values assumed clinically preferable. Let $\mathcal F_n=\sigma(\boldsymbol\delta_1,Y_1,\ldots,\boldsymbol\delta_n,Y_n)$ denote the information available after the first $n$ observations, throughout the paper treatment allocation is assumed to be sequential, with allocation probabilities $\Pr(\delta_{n+1,k}=1\mid\mathcal F_n)$ depending only on the information accrued up to that step (but not on the potential outcomes of the current subject). Moreover, for each $k$, let $N_{n,k}=\sum_{i=1}^n\delta_{ik}$ be the number of units assigned to treatment $k$ among the first $n$ subjects and we set $\boldsymbol{N}_n =(N_{n,1},\ldots,N_{n,K})^\top$; analogously, $\boldsymbol{\pi}_n=n^{-1}\boldsymbol{N}_n=(\pi_{n,1},\ldots,\pi_{n,K})^\top$ is the vector of allocation proportions to the arms, where $\boldsymbol{1}_K^\top\boldsymbol{N}_n=n$ and $\boldsymbol{1}_K^\top\boldsymbol{\pi}_n=1$ for every $n$ ($\boldsymbol{1}_K$ and $\boldsymbol{0}_K$ denote the $K$-dim vectors of ones and zeros, respectively; moreover, throughout the paper, $\E$ and $\var$ denote expectation and variance, $\1\{A\}$ is the indicator function of the event $A$, $\weak$ represents the weak convergence, while $\circ$ denotes function composition).

We assume that the potential-outcome vectors $(Y_{i1},\dots,Y_{iK})^\top$ ($i\ge1$) are independent and identically distributed (i.i.d.) as $(Y_1,\dots,Y_K)^\top$, with marginal CDFs $F_1,\dots,F_K$ defined on a common domain $\mathcal X\subseteq\mathbb R$, which contains the support of each treatment response. As an immediate consequence of the predictability of the allocation probabilities with respect to $\mathcal F_n$, for each treatment $k$ the sequence of observed responses on that arm has the same distributional structure as an i.i.d. sample from $F_k$ (see \cite{belisle_melfi}) and we refer to this property as independence-after-adaptive-allocation.

For any given pair of treatments $k,j\in\{1,\ldots,K\}$, the difference between the corresponding CDFs is closely connected to first-order stochastic dominance. Indeed, treatment $k$ dominates treatment $j$ if and only if $F_k(x)\le F_j(x)$ for all $x$, with strict inequality somewhere. In such a case $\E[g(Y_k)]\ge \E[g(Y_j)]$ for every increasing function $g(\cdot)$ for which the expectations exist, namely treatment $k$ is superior to $j$ under any transformation that preserves the ordering (the extension to $K>2$ is immediate). While first-order stochastic dominance guarantees a global superiority/inferiority of a treatment with respect to the others, if the CDFs cross over $\X$ the comparison can no longer be reduced to a universal ranking that is valid for all reasonable preference criteria. In this setting, neither treatment is uniformly preferable and the notion of treatment superiority/inferiority is not intrinsic to the CDFs alone, but depends on how the relevant features of the outcome distribution are aggregated. Thus, when stochastic dominance fails, comparison becomes inherently criterion-dependent: one treatment may be superior to the others in some regions of $\X$, while another could be superior elsewhere; therefore, the treatment ranking is not determined by the CDFs alone, but requires an explicit rule for aggregating the distributional performance across the entire outcome space.

Let $\mathcal G$ denote the space of CDFs on $\X$; this rule is encoded by a real-valued score functional $\Psi:\mathcal G\to\R$, which maps the CDF of each treatment into a scalar score by preserving the first-order stochastic dominance; since larger outcomes are clinically preferable, for every pair of treatments $k,j\in\{1,\ldots,K\}$, the score functional $\Psi$ should satisfy the following condition:
$$F_k(x)\le F_j(x)\ \ \forall x \quad\Longrightarrow\quad \Psi(F_k)\ge \Psi(F_j).$$
This perspective is particularly natural in clinical applications, where different therapeutic objectives may legitimately induce different treatment orderings. For example, given a clinically meaningful response threshold $\tau\in \X$, $\Psi^{\tau}(F_k)=1-F_k(\tau)=\Pr(Y_k>\tau)$ ranks the treatments according to their probability of exceeding $\tau$ (it is natural when the clinical question is framed in terms of remission or avoidance of severe disease). While this criterion provides a “local” evaluation of the CDFs around a critical level, other functionals could involve the whole behavior of $F_k$s, also adopting a nonnegative weight function $w:\X\to \mathbb{R}^+$, which makes explicit which regions of $\X$ are clinically emphasized. For instance, the functional
\begin{equation}\label{eq:linear-functional}
\Psi^L(F_k)=\int_\X \left[1-F_k(x)\right]w(x)dx
\end{equation}
is strictly related to the comparison based on the treatment effects; indeed, if $\X=\mathbb{R}^+$ and $w(x)\equiv 1$, then $\Psi^L(F_k)=\int_0^\infty \left[1-F_k(x)\right]dx=\E(Y_k)$ (provided that the expectations exist), recovering classical mean-based comparisons. Another functional very useful in practical applications is the quantile-based score $\Psi^\alpha(F_k)=F_k^{-1}(\alpha)=\inf\{x\in\mathcal X:F_k(x)\geq\alpha\}$, with $\alpha\in(0,1)$. For small values of $\alpha$, this score provides a direct assessment of the lower tail of the distribution and may be particularly relevant when the clinical objective is to protect patients against poor outcomes. Since larger outcomes are preferable, larger values of $\Psi^\alpha$ correspond to better performance. Obviously, $\Psi^\alpha$ preserves first-order stochastic dominance, since $F_k(x)\leq F_j(x)$ for all $x$ implies that $F_k^{-1}(\alpha)\geq F_j^{-1}(\alpha)$, namely $\Psi^\alpha(F_k)\geq \Psi^\alpha(F_j)$.
\begin{remark}
Along the same lines, other functional scores can be introduced to emphasize protection against outcomes above/below suitably chosen critical thresholds (that arise naturally when the priority is to reduce severe failures or disability). Moreover, if $u(\cdot)$ is an increasing utility function absolutely continuous on $\X$, we could define $\Psi^u(F_k)=\E\{u(Y_k)\}$. If $\X=\mathbb{R}^+$, then $\Psi^u(F_k)=\int_{0}^{\infty} u'(x)\bigl[1-F_k(x)\bigr]dx$ (up to an additive constant), where the derivative $u'$ acts as a local weight, indicating where improvements are most valuable.
\end{remark}
The following motivating examples illustrate how treatment rankings based on the complete response distributions (which may vary among the treatment arms) may differ substantially from those induced by conventional finite-dimensional summaries, and why this distinction may be relevant for RAR procedures.
\begin{example}\label{ex2.1}
Let $K=2$ with $Y_1\sim N(0.5;2.25)$ and $Y_2\sim N(0.3;0.25)$. Since $\E(Y_1)=0.5>0.3=\E(Y_2)$, a mean-based RAR rule favors the first treatment. However, assuming the quantile-based score $\Psi^{\alpha}$ with $\alpha=0.2$, then $\Psi^{0.2}(F_1)= -0.762$ and $\Psi^{0.2}(F_2)=-0.121$. Hence, if the clinical objective is protection of the lower tail, treatment 2 is clearly preferable. The same conclusion is obtained from the threshold score $\Psi^\tau$ at $\tau=0$, since $\Pr(Y_1>0)= 0.631$ and $\Pr(Y_2>0)= 0.726$. Thus, a distribution-based criterion may lead to a different treatment ranking from that induced by the means, since in the absence of stochastic dominance mean superiority need not coincide with superiority according to the clinically relevant feature of the response distribution.
\end{example}
\begin{example}\label{ex2.2}
Consider $K=3$ treatments with response distributions $Y_1\sim N(0;12)$, $Y_2\sim U(-6,6)$ and $Y_3$ follows a two-component Gaussian mixture $Y_3\sim 0.9N(1;3)+0.1N(-9;3)$. For every treatment $k$, $\E(Y_k)=0$ and $\var(Y_k)=12$, but their lower tails are markedly different. In particular, assuming the quantile-based score $\Psi^{\alpha}$ with $\alpha=0.025$, then $\Psi^{0.025}(F_1)= -6.79$, $\Psi^{0.025}(F_2)= -5.70$ and $\Psi^{0.025}(F_3)= -10.17 $. Thus, if protection against poor outcomes is the relevant clinical objective, treatment 2 is the best and treatment 3 is the worst one. The same ranking arises under the threshold score $\Psi^\tau$ with $\tau=-6$, since $\Pr(Y_1>-6)= 0.96$, $\Pr(Y_2>-6)=1$ and $\Pr(Y_3>-6)= 0.9$. Thus, treatment differences that cannot be captured by RAR procedures based only on the first two moments may be crucial in clinically relevant regions of the outcome distribution.
\end{example}
The previous example also highlights a more fundamental limitation of classical RAR procedures developed under a common parametric model across arms, for which treatment comparison is performed through a common finite-dimensional parameter (typically the mean or monotonic transformations of it). When the treatment distributions belong to different families, however, this parametric ordering is no longer naturally available. Even if a mean-based RAR rule were applied, differences among the treatment effects would generally provide no distributional justification for ranking the treatments, since the same ordering may coexist with markedly different tail behavior, quantiles, or crossing CDFs. In such settings, defining treatment performance directly through functionals of the response distributions provides a common comparison criterion without requiring the outcomes to share the same parametric family. Moreover, even if a conventional parametric RAR procedure could be adapted to the present framework by estimating the chosen distributional score through fitted parametric treatment distributions, model misspecification could still lead to incorrect treatment rankings and, consequently, to systematically distorted allocations.
\begin{example}\label{ex2.3}
Consider $K=2$ treatments with $Y_1\sim N(0;1)$ and $Y_2\sim 0.8N(1;1) +0.2N(-4;1)$ (where $\E(Y_2)=0$ and $\var(Y_2)=5$). Assuming the linear score $\Psi^L$ with $w(x)=\1\{x\in[-1,0]\}$ to emphasize performance over a relevant lower region of the outcome distribution, then $\Psi^L(F_1)= 0.684$ and $\Psi^L(F_2)= 0.740$, so treatment 2 is preferable according to the true response distributions. Suppose instead that the same score is estimated parametrically by fitting a Gaussian working model to each arm and evaluating $\Psi^L$ on the corresponding fitted distributions. While the Gaussian model is correctly specified for treatment 1, for treatment 2 the fitted distribution converges to $\widetilde F_2=N(0;5)$, yielding $\Psi^L(\widetilde F_2)=0.588<\Psi^L(F_1)=0.684$, whereas under the true response distributions $\Psi^L(F_2)= 0.740>\Psi^L(F_1)$. Hence, the working model (asymptotically) reverses the treatment ordering. Consequently, any RAR rule preserving the ordering of the estimated scores tends to favor treatment 1, although treatment 2 is actually preferable according to $\Psi^L$. The nonparametric approach avoids this phenomenon by estimating $\Psi^L(F_k)$ directly from the empirical CDFs, without imposing a common parametric family across treatment arms.
\end{example}
Therefore, when stochastic dominance fails, the choice of $\Psi$ (and, possibly, $w$) is a crucial component of the treatment comparison. A model-based criterion, usually employed in the RAR literature, is only a special case with $\Psi^L$ and uniform weighting over $\X=\mathbb{R}^+$. In many clinical settings, however, equal weighting is not appropriate: avoiding severe deterioration, achieving remission, or improving outcomes near a therapeutic threshold may be much more important than gains in less critical regions.

\section{The ethical NP-RAR procedure}\label{Sec3}
We first introduce an ethical NP-RAR procedure in which allocation probabilities are driven by the estimated distributional scores and are required to favor treatments exhibiting better clinical performance according to the pre-specified functional $\Psi$. After $n$ observations, for each treatment $k$, the empirical CDF based on the observed outcomes on that arm is $\widehat F_{n,k}(x)=N_{n,k}^{-1}\sum_{i=1}^n \delta_{ik}\,\1\{Y_i\le x\}$ for $x\in\X$, so the $K$-dim vector of empirical scores $\widehat{\boldsymbol{\Psi}}_n=(\Psi(\widehat F_{n,1}),\dots,\Psi(\widehat F_{n,K}))^\top$ summarizes the distributional evidence accumulated so far among the treatments and represents the natural estimator of $\boldsymbol{\Psi}=\bigl(\Psi(F_{1}),\ldots,\Psi(F_{K})\bigr)^\top$. Kernel smoothing is not required at this stage: since the ethical NP-RAR procedure is driven by functionals of the treatment CDFs, the empirical distribution functions provide natural estimators while avoiding bandwidth selection and allowing direct use of Glivenko-Cantelli and Donsker theory.

The procedure starts with a finite initialization phase of $n_0$ observations (by using, for instance, restricted randomization methods) in which
each treatment is assigned at least twice. Thereafter, the ethical NP-RAR design is defined by assigning the $(n+1)$th subject to treatment $k=1,\dots,K$ with probability
\begin{equation}\label{eq:direct-design}
\Pr(\delta_{n+1,k}=1\mid \F_{n})=\Lambda_k(\widehat{\boldsymbol{\Psi}}_{n}),\qquad n\geq n_0,
\end{equation}
where $\boldsymbol\Lambda=(\Lambda_1,\dots,\Lambda_K)^\top:\mathbb{R}^K\to\mathcal{S}_K$ is a continuous map taking values in the simplex $\mathcal{S}_K=\{(p_1,\dots,p_K):\;p_k>0,\sum_{k=1}^K p_k=1\}$. The interpretation is immediate: the trial learns each empirical treatment distribution, converts it into a scalar score, and then assigns the next patient according to a probability vector that favors treatments with better current scores.

Since larger values of the score correspond to better treatment performance, to formalize the ethical interpretation of this procedure we assume that:
\begin{itemize}
\item $\boldsymbol\Lambda$ preserves the ordering induced by the treatment scores, namely, for every $\boldsymbol{s}\in\mathbb{R}^K$ and every $k,j\in\{1,\ldots,K\}$, $s_k\geq s_j\Longrightarrow \Lambda_k(\boldsymbol{s})\geq\Lambda_j(\boldsymbol{s})$;
\item $\Lambda_k(\cdot)$ is nondecreasing in its $k$th argument, so that an improvement in terms of $s_k$ cannot reduce the allocation probability of treatment $k$ (with all the other scores held fixed).
\end{itemize}
When the treatments play symmetric roles, an additional natural requirement is permutation equivariance: for every permutation $\Pi$ of the treatment labels,
$\boldsymbol\Lambda(\Pi\boldsymbol{s})=\Pi\boldsymbol\Lambda(\boldsymbol{s})$ for each $\boldsymbol{s}\in\mathbb{R}^K$. Thus, every permutation of the scores induces the same permutation on $\boldsymbol\Lambda$, ensuring invariance with respect to treatment labeling (non-symmetric allocation rules may instead be appropriate when the treatments play structurally different roles, for instance when one arm is a pre-specified control, or when ethical/regulatory constraints differ across treatments).

For example, a natural choice is
\begin{equation}\label{allfuncK}
\Lambda_k(s_1,\dots,s_K)=\frac{\exp\{\eta s_k\}}{\sum_{j=1}^K \exp\{\eta s_j\}},\qquad \text{with }\eta>0, \quad  k=1,\dots,K,
\end{equation}
which is continuous, permutation-equivariant, nondecreasing in $s_k$ and preserves the ordering of the treatment scores. For $K=2$, $\delta_{i,1}=1-\delta_{i,2}=\delta_{i}$ for every $i\geq1 $ and $\Lambda_1=1-\Lambda_2$, so a natural special case is a map $\Lambda_1$ depending on the difference between the two scores:
\begin{equation}\label{allfuncK1}
\Pr(\delta_{n+1}=1\mid\F_n)=\Lambda_1(\widehat{\boldsymbol{\Psi}}_n)=f\left(\Psi(\widehat F_{n,1})-\Psi(\widehat F_{n,2})\right),
\end{equation}
with $f:\mathbb{R}\rightarrow (0,1)$ continuous, nondecreasing and such that $f(-t)=1-f(t)$. For instance, when $K=2$, $\boldsymbol\Lambda$ in (\ref{allfuncK}) simply reduces to the logistic function; indeed, $\Lambda_1(s_1,s_2)=\{1+\exp\{-\eta(s_1-s_2)\}\}^{-1}$, which belongs to the family in (\ref{allfuncK1}) with $f(t)=\{1+\exp(-\eta t)\}^{-1}$ (satisfying $f(-t)=1-f(t)$ for every $t$).

\section{Asymptotic properties of the ethical NP-RAR procedure}\label{Sec4}
By combining martingale and empirical-process arguments, this section states the basic asymptotic properties of the ethical NP-RAR design in (\ref{eq:direct-design}). In what follows we assume:
\begin{assumption}\label{ass:lambda}
For every $k$, $\liminf_{n\rightarrow\infty} \Lambda_k(\widehat{\boldsymbol{\Psi}}_n)>0$ a.s.
\end{assumption}
\begin{assumption}\label{ass:phi}
$\Psi$ is continuous at each $F_k$ with respect to the sup-norm.
\end{assumption}
Condition \ref{ass:lambda} guarantees that each treatment is sampled infinitely often a.s., which in turn ensures that all treatment-specific empirical distributions remain consistently estimable and provides the basic exploration condition needed for the empirical-process asymptotics; due to the continuity of $\Lambda_k(\cdot)$, this condition is directly satisfied if the range of the score $\Psi$ is bounded. Condition \ref{ass:phi} guarantees consistency of the functional scores.
\begin{theorem}\label{thm:gc}
Adopting the ethical NP-RAR procedure in (\ref{eq:direct-design}), under \ref{ass:lambda}-\ref{ass:phi}, as $n\to\infty$,
$$N_{n,k}\to\infty\qquad\text{and}\qquad \|\widehat F_{n,k}-F_k\|_\infty \to 0 \qquad a.s.\quad \text{for } k=1,\dots,K.$$
Thus, $\widehat{\boldsymbol{\Psi}}_n\to \boldsymbol{\Psi}$, $\boldsymbol{\Lambda}(\widehat{\boldsymbol{\Psi}}_n)\to \boldsymbol{\Lambda}(\boldsymbol{\Psi})$ and $\boldsymbol{\pi}_{n}\to \boldsymbol{\Lambda}(\boldsymbol{\Psi})$ a.s.
\end{theorem}
\begin{proof}
The proof is given in Appendix \ref{proofTHM1}.
\end{proof}
The next theorem provides the functional central limit theorems for the empirical CDFs and their functionals, as well as for the treatment allocation proportions. To derive the first-order asymptotics, some additional assumptions are required.
\begin{assumption}\label{ass:CLT2}
For every $k$, there exists a measurable function $\varphi_k:\mathcal X\to\mathbb R$ such that
\begin{equation}\label{linearizzazione}
\Psi(\widehat F_{n,k})-\Psi(F_k)=\frac{1}{N_{n,k}}\sum_{i=1}^n\delta_{ik}\varphi_k(Y_{ik})+r_{n,k},
\end{equation}
where $\E\{\varphi_k(Y_k)\}=0$, $\E\{|\varphi_k(Y_k)|^{2+\varepsilon}\}<\infty$ for some $\varepsilon>0$ and $\sum_{i=1}^n |r_{i,k}|=o_p(\sqrt{n})$.
\end{assumption}
Building on the consistency results of Theorem \ref{thm:gc}, the following theorem provides the corresponding first-order asymptotics. Hereafter, Hadamard differentiability is understood tangentially to a suitable subspace containing the limiting empirical process.
\begin{theorem}\label{thm:tec-clt}
Adopting the ethical NP-RAR rule in (\ref{eq:direct-design}), under \ref{ass:lambda}-\ref{ass:phi}, as $n\to\infty$,
\begin{equation}\label{CLTF}
\sqrt n\left(\widehat F_{n,1}-F_1,\ldots,\widehat F_{n,K}-F_K\right)^\top\weak \operatorname{diag}
\left\{\boldsymbol{\Lambda}(\boldsymbol{\Psi})\right\}^{-1/2}\Bigl(\mathbb G_1\circ F_1,\ldots, \mathbb G_K\circ F_K \Bigr)^\top
\end{equation}
in $\bigl\{\ell^\infty(\mathcal X)\bigr\}^K$, where $\mathbb G_1,\dots,\mathbb G_K$ are independent standard Brownian bridges. If, in addition, $\Psi$ is Hadamard differentiable at every $F_k$, with derivative $\dot\Psi_{F_k}$, then
\begin{equation}\label{CLTpsi}
\sqrt n(\widehat{\boldsymbol{\Psi}}_n- \boldsymbol{\Psi})\weak \boldsymbol Z,
\end{equation}
where $\boldsymbol Z=\Bigl(\dot\Psi_{F_1}\bigl[(\Lambda_1(\boldsymbol{\Psi}) )^{-1/2}\mathbb G_1\circ F_1\bigr],\ldots, \dot\Psi_{F_K}\bigl[(\Lambda_K(\boldsymbol{\Psi}) )^{-1/2}\mathbb G_K\circ F_K\bigr]\Bigr)^\top$.

Moreover, if $\boldsymbol\Lambda$ is differentiable at $\boldsymbol\Psi$, then
\begin{equation}\label{CLTlambda}
\sqrt n\{\boldsymbol{\Lambda}(\widehat{\boldsymbol{\Psi}}_n)-\boldsymbol{\Lambda}(\boldsymbol{\Psi})\} \weak \boldsymbol J\boldsymbol Z,
\end{equation}
where $\boldsymbol J$ denotes the Jacobian matrix of $\boldsymbol\Lambda$ evaluated at $\boldsymbol{\Psi}$, and if \ref{ass:CLT2} holds,
\begin{equation}\label{CLTpi}
\sqrt n \left\{\boldsymbol{\pi}_n-\boldsymbol{\Lambda}(\boldsymbol{\Psi})\right\}\Rightarrow N \left(\boldsymbol0_K;\operatorname{diag}
\left\{\boldsymbol{\Lambda}(\boldsymbol{\Psi})\right\}-\boldsymbol{\Lambda}(\boldsymbol{\Psi})\boldsymbol{\Lambda}(\boldsymbol{\Psi})^\top
+2\boldsymbol J\boldsymbol \Sigma \boldsymbol J^\top\right),
\end{equation}
where $\boldsymbol\Sigma= \operatorname{diag}\left(\E\{\varphi_k(Y_k)^2\}/\Lambda_k(\boldsymbol{\Psi})\right)_{k=1,\ldots,K}$.
\end{theorem}
\begin{proof}
The proof is given in Appendix \ref{proofTHM2}.
\end{proof}
From Theorem \ref{thm:tec-clt}, notice that:
\begin{itemize}
\item since $\boldsymbol{\pi}_{n}\to \boldsymbol{\Lambda}(\boldsymbol{\Psi})$ a.s., under the ethical NP-RAR procedure the map $\boldsymbol{\Lambda}$ determines both the allocation function and the asymptotic target; however, the empirical-process and score limits in Theorem \ref{thm:tec-clt} are not specific to the ethical NP-RAR. Indeed, by the independence-after-adaptive-allocation property introduced in Section \ref{Sec2}, for any allocation rule such that $\boldsymbol\pi_n\to\boldsymbol\rho\in\mathcal S_K$ a.s., limits \eqref{CLTF}-(\ref{CLTpsi}) continue to hold with $\boldsymbol{\Lambda}(\boldsymbol{\Psi})$ replaced by $\boldsymbol\rho$.
\item for each $k$, the functional convergence in (\ref{CLTF}) and (\ref{CLTpsi}) can be rewritten under the treatment-specific normalization $\sqrt{N_{n,k}}$ instead of $\sqrt{n}$; in such a case,
\begin{equation}\label{th2new}
\sqrt{N_{n,k}}\bigl(\widehat F_{n,k}-F_k\bigr)\Rightarrow \mathbb G_k\circ F_k\;\; \text{and} \;\; \sqrt{N_{n,k}}\bigl\{\Psi(\widehat F_{n,k})-\Psi(F_k)\bigr\}\Rightarrow \dot\Psi_{F_k}(\mathbb G_k\circ F_k).
\end{equation}
\item Let $\nu_k^2:= \var[\dot\Psi_{F_k}(\mathbb G_k\circ F_k)]>0$ for $k=1,\ldots,K$, from (\ref{CLTpsi}),
\begin{equation}\label{cltpsi2}
\sqrt n(\widehat{\boldsymbol{\Psi}}_n- \boldsymbol{\Psi})\weak N(\boldsymbol 0_K;\boldsymbol V(\boldsymbol\Lambda(\boldsymbol\Psi)))
\end{equation}
where
\begin{equation}\label{cltpsi2bis}
\boldsymbol V(\boldsymbol\Lambda(\boldsymbol\Psi))=\operatorname{diag}\left(\frac{\nu_k^2}{\Lambda_k(\boldsymbol\Psi)}\right)_{k=1,\ldots,K},
\end{equation}
due to the linearity of the Hadamard derivative and the independence of the limiting treatment-specific empirical processes. Analogously, from (\ref{CLTlambda}),
\begin{equation}\label{lambdanew}
\sqrt n\{\boldsymbol\Lambda(\widehat{\boldsymbol{\Psi}}_n)-\boldsymbol\Lambda(\boldsymbol\Psi)\}\Rightarrow N\left(\boldsymbol 0_K; \boldsymbol J\boldsymbol V(\boldsymbol\Lambda(\boldsymbol\Psi)) \boldsymbol J^\top\right).
\end{equation}
\item If $\Psi$ is Hadamard differentiable at every $F_k$, \ref{ass:CLT2} holds and, in addition,
\begin{equation}\label{eq:strongremainder}
r_{n,k}=o_{a.s}\left(N_{n,k}^{-1/2}\right),\qquad \text{for } k=1,\ldots,K,
\end{equation}
then $\nu_k^2=\E\{\varphi_k(Y_k)^2\}$ and hence $\boldsymbol\Sigma= \boldsymbol V(\boldsymbol\Lambda(\boldsymbol\Psi))$. Indeed, from (\ref{th2new}),
$\sqrt{N_{n,k}}\{\Psi(\widehat F_{n,k})-\Psi(F_k)\}\Rightarrow N(0;\nu_k^2)$. However, \ref{ass:CLT2}
together with (\ref{eq:strongremainder}) yields
$$\sqrt{N_{n,k}}\left\{\Psi(\widehat F_{n,k})-\Psi(F_k)\right\}=\frac{1}{\sqrt{N_{n,k}}}\sum_{i=1}^n\delta_{ik}\varphi_k(Y_{ik})+o_p(1),$$
whose limiting distribution is centered normal with variance $\E\{\varphi_k(Y_k)^2\}$. Due to the uniqueness of the limiting distribution, $\nu_k^2=\E\{\varphi_k(Y_k)^2\}$.
\item No moment assumptions on the treatment outcomes are required for the empirical-CDF limit in (\ref{CLTF}), since the underlying indicator functions are uniformly bounded. Specific integrability or moment conditions should be imposed only when required by a particular choice of $\Psi$. Clearly, scores involving suprema, positive parts or signs may fail to be fully Hadamard differentiable, but they can often be handled through directional differentiability and delta methods.
\end{itemize}
\begin{remark}
The asymptotic variance of the allocation proportions in (\ref{CLTpi}) has a natural decomposition which provides a direct interpretation of the two distinct sources of variability of the design; indeed,
$\boldsymbol Q:=\operatorname{diag}\{\boldsymbol\Lambda(\boldsymbol\Psi)\}-\boldsymbol\Lambda(\boldsymbol\Psi)\boldsymbol\Lambda(\boldsymbol\Psi)^\top$ represents the intrinsic variability induced by randomization, while $2\boldsymbol J\boldsymbol\Sigma \boldsymbol J^\top$ reflects the additional variability generated by the adaptation (via the estimated scores). This form is a consequence of the nature of the proposed NP-RAR rule: at each step, the current estimate $\boldsymbol\Lambda(\widehat{\boldsymbol{\Psi}}_n)$ of the limiting allocation $\boldsymbol\Lambda(\boldsymbol\Psi)$ is used directly as the randomization probability in \eqref{eq:direct-design}, without any feedback from the current allocation proportions. This observation suggests separating the estimation of the target from the mechanism used to track it. In particular, the suggested nonparametric framework can be embedded into the ERADE-type mechanism \citep{Hu09} to reduce the first-order variability of the allocation proportions. The resulting extension is considered in Section \ref{Sec6}.
\end{remark}

\subsection{Examples}
We now illustrate the general results of Theorems \ref{thm:gc} and \ref{thm:tec-clt} through some relevant special cases of the ethical NP-RAR procedure. The first three examples concern different meaningful choices of the score functional $\Psi$. For each of them, we verify the required regularity conditions and derive the corresponding asymptotic distributions of the empirical scores, assignment probabilities and allocation proportions. Example \ref{ex_lambda} instead focuses on the role played by $\boldsymbol\Lambda$.

\begin{example}\label{ex_tau}
Consider the threshold functional $\Psi^\tau(F_k)=1-F_k(\tau)$. Let $v\in \ell^\infty(\mathcal X)$ be a perturbation function and consider
$F_k+tv$ (with $t\downarrow 0$). Then $\Psi^\tau(F_k+tv)=1-(F_k+tv)(\tau)=1-F_k(\tau)-tv(\tau)$; hence,
$$\frac{\Psi^\tau(F_k+tv)-\Psi^\tau(F_k)}{t}=-v(\tau),$$
so the Hadamard derivative of $\Psi^\tau$ at $F_k$ in direction $v$ is simply $\dot\Psi_{F_k}^{\tau}(v)=-v(\tau)$.

Thus, $\sqrt{N_{n,k}}\bigl(\Psi^\tau(\widehat F_{n,k})-\Psi^\tau(F_k)\bigr) \Rightarrow -\mathbb G_k(F_k(\tau))$. Since $\mathbb G_k$ is a standard Brownian bridge, for every fixed $u\in[0,1]$, $ \mathbb G_k(u)\sim N(0;u[1-u])$ and therefore, from (\ref{th2new}),
$$\sqrt{N_{n,k}}\bigl(\Psi^\tau(\widehat F_{n,k})-\Psi^\tau(F_k)\bigr) \Rightarrow N\left(0;F_k(\tau)\bigl[1-F_k(\tau)\bigr]\right).$$
If, in addition, $\boldsymbol\Lambda$ is differentiable at $\boldsymbol\Psi$, then (\ref{lambdanew}) holds with
$$\boldsymbol V(\boldsymbol\Lambda(\boldsymbol\Psi))=\operatorname{diag}\left(\frac{F_k(\tau)[1-F_k(\tau)]}{\Lambda_k(\boldsymbol\Psi)}\right)_{k=1,\ldots,K}.$$
Moreover, \ref{ass:CLT2} is satisfied; indeed, let $\varphi_{\tau,k}(y)=F_k(\tau)-\1\{y\le\tau\}$, then
$$\Psi^\tau(\widehat F_{n,k})-\Psi^\tau(F_k)=-\{\widehat F_{n,k}(\tau)-F_k(\tau)\}=\frac1{N_{n,k}}\sum_{i=1}^n\delta_{ik}\varphi_{\tau,k}(Y_{ik})$$
(i.e. $r_{n,k}=0$ for every $k$ and $n$). Since $\E\{\varphi_{\tau,k}(Y_k)\}=0$ and $\E\{\varphi_{\tau,k}(Y_k)^2\}=F_k(\tau)[1-F_k(\tau)]$, then (\ref{CLTpi}) holds with
$\boldsymbol\Sigma=\boldsymbol V(\boldsymbol\Lambda(\boldsymbol\Psi))$, namely
$$\sqrt n \left\{\boldsymbol{\pi}_n-\boldsymbol{\Lambda}(\boldsymbol{\Psi})\right\}\Rightarrow N \left(\boldsymbol0_K,\operatorname{diag}\left\{\boldsymbol{\Lambda}(\boldsymbol{\Psi})\right\}-\boldsymbol{\Lambda}(\boldsymbol{\Psi})\boldsymbol{\Lambda}(\boldsymbol{\Psi})^\top
+2\boldsymbol J\boldsymbol \Sigma \boldsymbol J^\top\right).$$
\end{example}

\begin{example}\label{ex_L}
Consider $\Psi^L(F_k)$ in (\ref{eq:linear-functional}) with $\X=[0,1]$ and $w(x)\equiv 1$, namely $\Psi^L(F_k)=\E(Y_k)<\infty$. In such a case $\Psi^L$ is Lipschitz continuous with respect to the sup-norm and its Hadamard derivative at $F_k$ in direction $v$ is $\dot\Psi_{F_k}^{L}(v)=-\int_0^1 v(x)dx$; therefore,
$$\sqrt{N_{n,k}}\Bigl(\Psi^L(\widehat F_{n,k})-\Psi^L(F_k)\Bigr)\Rightarrow -\int_0^1 \mathbb G_k(F_k(x))dx.$$
Thus, the limit is a centered Gaussian random variable with variance
$$\var\left(\int_0^1\mathbb G_k(F_k(x))dx\right)=\int_0^1\int_0^1 \operatorname{Cov}\bigl(\mathbb G_k(F_k(x)),\mathbb G_k(F_k(y))\bigr)dxdy.$$
Since $\mathbb G_k$ is a standard Brownian bridge, $\operatorname{Cov}(\mathbb G_k(u),\mathbb G_k(\tilde{u}))=u\wedge \tilde{u}-u\tilde{u}$, where $u\wedge \tilde{u}=\min\{u,\tilde{u}\}$; recalling that $F_k$ is nondecreasing, the asymptotic variance is
$$\int_0^1\int_0^1 \Bigl[F_k(x\wedge y)-F_k(x)F_k(y)\Bigr]dxdy=\var(Y_k)$$
(which is finite since $\X=[0,1]$) and, from (\ref{th2new}), $\sqrt{N_{n,k}}(\Psi^L(\widehat F_{n,k})-\Psi^L(F_k))\Rightarrow N(0;\var(Y_k))$.
Notice that, in such a case, $\Psi^L(F_k)=\E(Y_k)$ so the empirical score coincides with the treatment-specific sample mean; indeed, if $\mathcal X=[0,1]$ and $w(x)\equiv 1$, then
$$\Psi^L(\widehat F_{n,k})=\int_0^1 \frac{1}{N_{n,k}}\sum_{i=1}^n \delta_{ik}\1\{Y_i>x\}dx= \frac{1}{N_{n,k}}\sum_{i=1}^n \delta_{ik}\int_0^1 \1\{Y_i>x\}dx=\frac{1}{N_{n,k}}\sum_{i=1}^n \delta_{ik}Y_i$$
and Theorem \ref{thm:tec-clt} recovers as a special case the classical central limit theorem for treatment-specific sample means under RAR procedures.

If, in addition, $\boldsymbol\Lambda$ is differentiable at $\boldsymbol\Psi$, the asymptotic normality in (\ref{lambdanew}) holds with
$\boldsymbol V(\boldsymbol\Lambda(\boldsymbol\Psi))=\operatorname{diag}\left(\var(Y_k)/\Lambda_k(\boldsymbol\Psi)\right)_{k=1,\ldots,K}$. Furthermore, \ref{ass:CLT2} holds exactly; indeed, $\Psi^L(\widehat F_{n,k})-\Psi^L(F_k)=N_{n,k}^{-1}\sum_{i=1}^n\delta_{ik}\{Y_{ik}-\E(Y_k)\}$, so that $\varphi_{L,k}(y)=y-\E(Y_k)$ and $r_{n,k}=0$ for every $k$ and $n$. Since $\mathcal X=[0,1]$, $\varphi_{L,k}$ is bounded and the $(2+\varepsilon)$ moment condition of \ref{ass:CLT2} is satisfied with $\E\{\varphi_{L,k}(Y_k)\}=0$ and $\E\{\varphi_{L,k}(Y_k)^2\}=\var(Y_k)$. Thus, (\ref{CLTpi}) holds with $\boldsymbol\Sigma=\boldsymbol V(\boldsymbol\Lambda(\boldsymbol\Psi))$.
\end{example}

\begin{example}\label{ex_q}
Consider the quantile-based score $\Psi^\alpha(F_k)$ and assume that $F_k$ admits a continuously differentiable density $f_k$ in a neighborhood of $\Psi^\alpha(F_k)$, with $f_k(\Psi^\alpha(F_k))>0$ and $f_k'$ bounded in that neighborhood. Thus, $\Psi^\alpha$ is Hadamard differentiable at $F_k$ with
$\dot\Psi_{F_k}^{\alpha}(v)=-v(\Psi^\alpha(F_k))/f_k(\Psi^\alpha(F_k))$. Since $\Psi^{\alpha}(\widehat F_{n,k})=\widehat F_{n,k}^{-1}(\alpha)$ is the empirical $\alpha$-quantile, Theorem \ref{thm:tec-clt} yields
$$\sqrt{N_{n,k}}\left\{\Psi^{\alpha}(\widehat F_{n,k})-\Psi^{\alpha}(F_k)\right\}\Rightarrow -\frac{\mathbb{G}_k\bigl(F_k(\Psi^\alpha(F_k))\bigr)}{f_k(\Psi^\alpha(F_k))}.$$
Since $F_k$ is continuous at $\Psi^\alpha(F_k)$, then $F_k(\Psi^\alpha(F_k))=F_k( F_k^{-1}(\alpha))=\alpha$ and
$$\sqrt{N_{n,k}}\left(\Psi^{\alpha}(\widehat F_{n,k})-\Psi^\alpha(F_k)\right)\Rightarrow N\left(0, \frac{\alpha(1-\alpha)}{\left[f_k(\Psi^\alpha(F_k))\right]^2}\right).$$
Thus, the proposed framework recovers, as a special case, the classical asymptotic distribution of empirical quantiles, while allowing the quantile itself to drive the response-adaptive allocation. This criterion depends directly on a specific feature of the treatment distribution and cannot in general be represented as the expectation of a transformation of the outcome.

Moreover, if $\boldsymbol\Lambda$ is differentiable at $\boldsymbol\Psi$, then (\ref{lambdanew}) holds with
$$\boldsymbol V(\boldsymbol\Lambda(\boldsymbol\Psi))=
\alpha(1-\alpha)\operatorname{diag}\left(\Lambda_k(\boldsymbol\Psi)\left[f_k\left(\Psi^\alpha(F_k)\right)\right]^2\right)_{k=1,\ldots,K}^{-1}.$$
As regards the asymptotic distribution of the allocation proportions, under the above regularity conditions, by applying the Bahadur-Kiefer representation (see, e.g., \cite{Bahadur1966,Kiefer1967}), decomposition (\ref{linearizzazione}) of Condition \ref{ass:CLT2} holds with
$$\varphi_{\alpha,k}(y)=\frac{\alpha-\1\{y\le\Psi^\alpha(F_k)\}}{f_k(\Psi^\alpha(F_k))}$$
and $r_{n,k}=O_{\mathrm{a.s.}}[N_{n,k}^{-3/4}(\log N_{n,k})^{1/2}(\log\log N_{n,k})^{1/4}]$. Thus, condition (\ref{eq:strongremainder}) is satisfied
since $N_{n,k}^{-3/4}(\log N_{n,k})^{1/2}(\log\log N_{n,k})^{1/4}=o(N_{n,k}^{-1/2})$ and, recalling that $\pi_{n,k}\to\Lambda_k(\boldsymbol\Psi)>0$ a.s., then
$n^{-1/2}\sum_{m=1}^n |r_{m,k}| \rightarrow 0$ a.s. Moreover, $\E\{\varphi_{\alpha,k}(Y_k)\}=0$ and $\E\{\varphi_{\alpha,k}(Y_k)^2\}=\alpha(1-\alpha)/\left[f_k\{\Psi^\alpha(F_k)\}\right]^2$, \ref{ass:CLT2} is satisfied by observing that $\varphi_{\alpha,k}$ is bounded (since $f_k(\Psi^\alpha(F_k))>0$). Therefore, the asymptotic normality in (\ref{CLTpi}) holds with $\boldsymbol\Sigma=\boldsymbol V(\boldsymbol\Lambda(\boldsymbol\Psi))$.
\end{example}
The following example is focused on the role played by $\boldsymbol\Lambda$ in determining the asymptotic behavior of the ethical NP-RAR procedure, while leaving the score $\Psi$ otherwise unspecified.
\begin{example}\label{ex_lambda}
Consider the allocation rule in (\ref{allfuncK}). For every $k,j=1,\ldots,K$,
$$\frac{\partial\Lambda_k}{\partial s_j}=
\begin{cases}
\eta\Lambda_k(1-\Lambda_k), & j=k,\\[1mm]
-\eta\Lambda_k\Lambda_j, & j\neq k.
\end{cases}$$
Thus, $\boldsymbol J = \eta \left[\operatorname{diag}\{\boldsymbol\Lambda(\boldsymbol\Psi)\}-\boldsymbol\Lambda(\boldsymbol\Psi)\boldsymbol\Lambda(\boldsymbol\Psi)^\top\right]= \eta \boldsymbol Q$. Hence, from (\ref{CLTlambda}) and (\ref{CLTpi}),
$$\sqrt n \left\{\boldsymbol\Lambda(\widehat{\boldsymbol{\Psi}}_n)-\boldsymbol\Lambda(\boldsymbol\Psi)\right\}\Rightarrow N\left(\boldsymbol 0_K,\eta^2\boldsymbol Q \boldsymbol V(\boldsymbol\Lambda(\boldsymbol\Psi)) \boldsymbol Q^\top\right)$$ and
$\sqrt n\left\{\boldsymbol\pi_n-\boldsymbol\Lambda(\boldsymbol\Psi)\right\}\Rightarrow N\left(\boldsymbol 0_K,\boldsymbol Q+2\eta^2\boldsymbol Q \boldsymbol \Sigma \boldsymbol Q^\top\right)$. Thus, $\eta$ controls the sensitivity of the allocation probabilities to differences among the treatment scores and thereby affects the additional asymptotic variability generated by adaptation.

For $K=2$ treatments, $\Lambda_1(\boldsymbol\Psi)=\{1+\exp\{-\eta(\Psi(F_1)-\Psi(F_2))\}\}^{-1}=1-\Lambda_2(\boldsymbol\Psi)$. Thus, $\boldsymbol Q=\Lambda_1(\boldsymbol\Psi)(1-\Lambda_1(\boldsymbol\Psi))\boldsymbol d\boldsymbol d^\top$, where $\boldsymbol d=(1,-1)^\top$ and $\boldsymbol J=\eta \Lambda_1(\boldsymbol\Psi)(1-\Lambda_1(\boldsymbol\Psi)) \boldsymbol d\boldsymbol d^\top$. Thus, $\sqrt n\left(\pi_{n,1}-\Lambda_1(\boldsymbol\Psi)\right)$ is asymptotically distributed as a centered normal with variance
$$ \Lambda_1(\boldsymbol\Psi)(1-\Lambda_1(\boldsymbol\Psi))\left\{1 + 2\eta^2 \left[\E\{\varphi_1(Y_1)^2\}(1-\Lambda_1(\boldsymbol{\Psi}))+\E\{\varphi_2(Y_2)^2\}\Lambda_1(\boldsymbol{\Psi})\right]\right\}$$
(the corresponding result for the second treatment follows from $\pi_{n,2}=1-\pi_{n,1}$).
\end{example}

\section{Inferentially optimal distribution-based targets}\label{Sec5}
The NP-RAR rule considered in Sections \ref{Sec3}-\ref{Sec4} is primarily motivated by ethical considerations, since the map $\boldsymbol{\Lambda}$ (which plays a dual role by determining the allocation probabilities and, at the same time, the asymptotic target) favors treatments exhibiting better performance according to the chosen score functional $\Psi$. A different perspective arises when the treatment allocation proportions are selected with the primary objective of improving inferential precision. The asymptotic results of Theorem \ref{thm:tec-clt} provide a natural basis for defining such inferentially optimal distribution-based targets.

\subsection{Optimal targets for score estimation}
Consider a generic allocation rule such that $\boldsymbol{\pi}_n\to\boldsymbol{\rho}=(\rho_1,\ldots,\rho_K)^\top\in\mathcal S_K$ a.s., from (\ref{cltpsi2}) and (\ref{cltpsi2bis}), the asymptotic variance of $\widehat{\boldsymbol{\Psi}}_n$ is
$\boldsymbol{V}(\boldsymbol{\rho})=\operatorname{diag}\left(\nu_k^2/\rho_k\right)_{k=1,\ldots,K}$,
where $\nu_k^2 =\var[\dot\Psi_{F_k}(\mathbb G_k\circ F_k)]>0$ for $k=1,\ldots,K$. Thus, classical optimality criteria can be directly formulated over the
simplex $\mathcal S_K$. While $D$-optimality minimizes the volume of the concentration ellipsoid of $\boldsymbol{\Psi}$ (i.e., the determinant of $\boldsymbol{V}(\boldsymbol{\rho})$), $A$-optimality minimizes the total marginal variability of the estimated scores (through the trace of the asymptotic variance), whereas $E$-optimality minimizes the largest principal variance, thus controlling the worst-direction uncertainty. The following theorem provides the optimal designs for estimating $\boldsymbol{\Psi}$.
\begin{theorem}\label{th:optimal_scores}
The $D$-optimal design for the estimation of the score $\boldsymbol{\Psi}$ is balanced, while the $A$- and $E$-optimal targets $\boldsymbol\rho^A$ and $\boldsymbol\rho^E$ are respectively given by
$$\rho_k^A=\frac{\nu_k}{\sum_{j=1}^K\nu_j} \qquad \text{and} \qquad \rho_k^E =\frac{\nu_k^2}{\sum_{j=1}^K\nu_j^2}, \qquad k=1,\ldots,K.$$
\end{theorem}
\begin{proof}
The proof is given in Appendix \ref{proofTHM4}.
\end{proof}
Theorem \ref{th:optimal_scores} shows that the inferentially optimal allocation depends on the local variability of the chosen score, except under $D$-optimality,
for which the balanced design is always optimal. Thus, when the inferential criterion concerns the minimization of the volume of the confidence region for the score vector, balance is optimal independently of the treatment distributions. Recalling that, under the Hadamard differentiability assumptions, \ref{ass:CLT2} and (\ref{eq:strongremainder}), $\nu_k^2=\E\{\varphi_k(Y_k)^2\}$, then the $A$- and $E$-optimal targets can equivalently be expressed in terms of the second moments of the influence functions appearing in the asymptotic linear representation (\ref{linearizzazione}). Contrary to $D$-optimality, these allocations depend on the functionals of the treatment distributions, which leads to a broader class of local optimality problems, in which the optimal design depends on distributional functionals rather than on unknown finite-dimensional model parameters.

For instance, the $A$- and $E$-optimal targets corresponding to the score functionals considered in Examples \ref{ex_tau}-\ref{ex_q} follow immediately. Assuming the score $\Psi^\tau$,
\begin{equation}\label{Aotttau}
\rho_{k,\tau}^{A}=\frac{\sqrt{F_k(\tau)\{1-F_k(\tau)\}}}{\sum_{j=1}^K\sqrt{F_j(\tau)\{1-F_j(\tau)\}}}
\quad \text{and} \quad \rho_{k,\tau}^{E} = \frac{F_k(\tau)\{1-F_k(\tau)\}}{\sum_{j=1}^K F_j(\tau)\{1-F_j(\tau)\}},
\end{equation}
so that the $A$- and $E$-optimal allocations assign a larger fraction of observations to treatments for which $F_k(\tau)$ is closer to
$1/2$, namely those for which the probability of exceeding the clinically relevant threshold is estimated with larger variability.
While for the score $\Psi^L$ of Example \ref{ex_L}, where $\Psi^L(F_k)=\E(Y_k)$,
$$\rho_{k,L}^{A}=\frac{ \sqrt{\var(Y_k)}}{\sum_{j=1}^K\sqrt{\var(Y_j)}}\qquad \text{and} \qquad  \rho_{k,L}^{E}=\frac{\var(Y_k)}{\sum_{j=1}^K\var(Y_j)},$$
so that the $A$-optimal target coincides with the classical Neyman allocation.
Finally, for the quantile score $\Psi^\alpha(F_k)=F_k^{-1}(\alpha)$,
\begin{equation}\label{Aottalpha}
\rho_{k,\alpha}^{A}=\frac{\bigl[f_k\{\Psi^\alpha(F_k)\}\bigr]^{-1}}{ \sum_{j=1}^K \bigl[f_j\{\Psi^\alpha(F_j)\}\bigr]^{-1}}\qquad \text{and} \qquad \rho_{k,\alpha}^{E}=\frac{\bigl[f_k\{\Psi^\alpha(F_k)\}\bigr]^{-2}}{  \sum_{j=1}^K\bigl[f_j\{\Psi^\alpha(F_j)\}\bigr]^{-2}},
\end{equation}
so that treatments having smaller density at the $\alpha$-quantile receive a larger allocation, reflecting the greater local difficulty in estimating the corresponding quantile.

\subsection{Optimal targets for distribution estimation}
The same optimal-design perspective can be applied when the inferential objective is the joint estimation of the complete treatment distributions rather than of pre-specified scores. From the functional central limit theorem in (\ref{CLTF}), under generic limiting allocation proportions $\boldsymbol{\rho}$, the asymptotic empirical process associated with treatment $k$ has covariance kernel $\rho_k^{-1}\left[F_k(x\wedge y)-F_k(x)F_k(y)\right]$. Thus, functional analogues of classical optimal-design criteria can be defined directly in terms of the covariance structure of the limiting empirical processes (see, e.g., \cite{Aletti2016,Cardot2011}). Let $\mathcal X$ be a possibly unbounded interval containing the supports of all treatment distributions. Under the $L^2(\mathcal{X},dx)$ topology, define $0<\mathcal{A}_k:=\int_{\mathcal{X}}F_k(x)\{1-F_k(x)\}dx<\infty$ for each $k$; the trace of the covariance operator corresponding to treatment $k$ is then $\mathcal{A}_k/\rho_k$ and a natural functional analogue of $A$-optimality is obtained by minimizing the total integrated asymptotic variance:
$$\min_{\boldsymbol{\rho}\in\mathcal{S}_K}\sum_{k=1}^K\frac{\mathcal{A}_k}{\rho_k}.$$
Adopting the same argument underlying the $A$-optimal allocation in Theorem \ref{th:optimal_scores} (with $\nu_k^2$ replaced by $\mathcal{A}_k$), the functional $A$-optimal target for the estimation of the CDFs is $\tilde{\boldsymbol\rho}^{A}$ with
\begin{equation}\label{funcAottalpha}
\tilde{\rho}_k^{A}=\frac{\sqrt{\mathcal{A}_k}}{\sum_{j=1}^K\sqrt{\mathcal{A}_j}}, \qquad  \text{for }  k=1,\ldots,K,
\end{equation}
which allocates a larger fraction of observations to treatments whose CDFs exhibit a larger integrated variance. If $\E|Y_k|<\infty$, then $\mathcal{A}_k=\frac{1}{2}\E|Y_k-Y_k'|$, where $Y_k'$ is an independent copy of $Y_k$. Thus, $\tilde{\boldsymbol\rho}^{A}$ is directly related to the intrinsic dispersion of each treatment distribution, measured through its Gini mean difference.

Alternatively, under the $\ell^\infty(\mathcal X)$ topology, the natural functional analogue of $E$-optimality is obtained by controlling the largest pointwise asymptotic variance. By letting $\mathcal{B}_k=\sup_{x\in\mathcal{X}}F_k(x)\{1-F_k(x)\}$, the corresponding mini-max criterion is $$\min_{\boldsymbol{\rho}\in\mathcal{S}_K}\max_{k=1,\ldots,K}\frac{\mathcal{B}_k}{\rho_k},$$
whose solution is the functional $E$-optimal target $\tilde{\boldsymbol\rho}^{E}$, where $\tilde{\rho}_k^{E}=\mathcal{B}_k/(\sum_{j=1}^K \mathcal{B}_j)$ for $k=1,\ldots,K$ (provided that the treatment distributions are non-degenerate). Clearly, if each $F_k$ is continuous, then $\mathcal{B}_k=1/4$ for every $k$ and $\tilde{\boldsymbol\rho}^{E}$ reduces to the balanced allocation (nontrivial optimal targets may arise for discontinuous treatment distributions or, alternatively, when precision is required over a pre-specified region $\mathcal X_0\subset\X$).
\begin{example}\label{ex:optimal_normal_cdf}
Suppose that $Y_k\sim N(\mu_k;\sigma_k^2)$ for $k=1,\ldots,K$. Due to the continuity of the normal distribution $\tilde{\boldsymbol\rho}^{E}=K^{-1}\boldsymbol{1}_K$. Moreover, since $Y_k-Y_k'\sim N(0;2\sigma_k^2)$, then
$\E|Y_k-Y_k'|=2\sigma_k/\sqrt{\pi}$ and hence $\mathcal{A}_k=\sigma_k/\sqrt{\pi}$. Therefore, when the inferential objective is estimation of the treatment CDFs, the functional $A$-optimal allocation $\tilde{\boldsymbol\rho}^{A}$ is given by
\begin{equation}\label{funcAopt}
\widetilde\rho_k^A =\frac{\sqrt{\sigma_k}}{\sum_{j=1}^K\sqrt{\sigma_j}}, \qquad \text{for }k=1,\ldots,K,
\end{equation}
which is proportional to the square root of the treatment standard deviations. Accordingly, even for the same collection of CDFs, the optimal target depends on the inferential goals.
\end{example}

\section{A general NP-ERADE for distribution-based targets}\label{Sec6}
The previous sections suggest separating two distinct components of a RAR procedure: the definition of the desired target and the randomization mechanism used to track it. In the ethical NP-RAR procedure of Section \ref{Sec3}, these two components coincide, since $\boldsymbol{\Lambda}(\widehat{\boldsymbol{\Psi}}_n)$ is used directly as the
randomization probability and $\boldsymbol{\Lambda}(\boldsymbol{\Psi})$ represents the corresponding asymptotic target. The inferential criteria introduced in Section \ref{Sec5}, however, lead to distribution-dependent targets that need not be induced by a pre-specified score. To accommodate both situations within a common framework, we now regard a generic target $\boldsymbol{\rho}$ as a functional of the treatment CDFs, namely $$\boldsymbol{\rho}=\boldsymbol{\rho}(F_1,\ldots,F_K)=\bigl(\rho_1(F_1,\ldots,F_K),\ldots,\rho_K(F_1,\ldots,F_K)\bigr)^\top\in\mathcal S_K,$$
and let $\widehat{\boldsymbol{\rho}}_n =(\widehat\rho_{n,1},\ldots,\widehat\rho_{n,K})^\top$, with $\widehat\rho_{n,k}=\rho_{k}(\widehat F_{n,1},\ldots,\widehat F_{n,K})$, be an $\mathcal F_n$-measurable estimator of $\boldsymbol{\rho}$ based on the accumulating data after $n$ steps (also allowing regularized versions of the plug-in estimator, provided that the regularization preserves the conditions stated below). For instance, the ethical target of Sections \ref{Sec3}-\ref{Sec4} corresponds to $\boldsymbol{\rho}(F_1,\ldots,F_K)=\boldsymbol{\Lambda}(\boldsymbol{\Psi})$, whereas the distributional targets in Section \ref{Sec5} provide different choices of $\boldsymbol{\rho}$.

In order to asymptotically approach such distribution-based targets, we now introduce a suitable nonparametric extension of the multi-arm ERADE recently proposed by Alkhnefr et al. \cite{Hu2025}, hereafter referred to as NP-ERADE. Unlike the original ERADE framework, which is formulated in terms of finite-dimensional treatment-specific parameters within a parametric model, the present NP-ERADE allows the allocation target to depend directly on the treatment distributions without requiring the responses to belong to a common parametric family. At each step $n$, let
$\mathcal O_n =\{j=1,\ldots,K:\pi_{n,j}>\widehat\rho_{n,j}\}$ and $\mathcal I_n=\{j=1,\ldots,K:\pi_{n,j}<\widehat\rho_{n,j}\}$; the NP-ERADE assigns the $(n+1)$th patient to treatment $k$ with probability
\begin{equation}\label{NPERADE}
\Pr(\delta_{n+1,k}=1\mid\mathcal F_n)=
\begin{cases}
\gamma\widehat\rho_{n,k}, & \pi_{n,k}>\widehat\rho_{n,k},\\[1mm]
\widehat\rho_{n,k},& \pi_{n,k}=\widehat\rho_{n,k},\\[1mm]
\displaystyle
\widehat\rho_{n,k}+\frac{1-\gamma}{|\mathcal I_n|}\sum_{j\in\mathcal O_n}\widehat\rho_{n,j},& \pi_{n,k}<\widehat\rho_{n,k},
\end{cases}
\end{equation}
where $\gamma\in(0,1)$ is a tuning parameter. In such a way, over-represented treatments (relative to the corresponding estimated target) receive a reduced allocation probability, while the corresponding probability mass is redistributed among under-represented arms. The target and the randomization rule now play different roles. In particular, the allocation probabilities in (\ref{NPERADE}) need not preserve the ordering of the treatment scores, since their purpose is to correct deviations of the current allocation proportions from the desired target. The ethical ordering requirement is instead encoded in the target itself whenever $\boldsymbol{\rho}=\boldsymbol{\Lambda}(\boldsymbol{\Psi})$. We now introduce the conditions needed to establish consistency and first-order asymptotics of the NP-ERADE.
\begin{assumption}\label{assERADE1}
For every $k$, $\liminf_{n\to\infty}\widehat\rho_{n,k}>0$ a.s. and the functional target $\boldsymbol{\rho}$ is continuous at $(F_1,\ldots,F_K)$ with respect to the product sup-norm.
\end{assumption}
\begin{assumption}\label{assERADE2}
For every $k$, there exist measurable functions $\boldsymbol h_{\rho,k}:\mathcal X\to\mathbb R^K$ such that
\begin{equation}\label{rho-linear}
\widehat{\boldsymbol{\rho}}_n-\boldsymbol{\rho}=\sum_{k=1}^K\frac{1}{N_{n,k}}\sum_{i=1}^n\delta_{ik}\boldsymbol h_{\rho,k}(Y_{ik})+\boldsymbol r_n^\rho,
\end{equation}
where $\E\{\boldsymbol h_{\rho,k}(Y_k)\}=\boldsymbol 0_K$, $\E\{\|\boldsymbol h_{\rho,k}(Y_k)\|^{2+\epsilon}\}<\infty$ for some $\epsilon>0$ and $\|\boldsymbol r_n^\rho\|=o_{\mathrm{a.s.}}(n^{-1/2})$.
\end{assumption}
\begin{assumption}\label{assERADE3}
For every $k$, $F_k$ belongs to an unrestricted nonparametric model with tangent space $L_0^2(F_k)$. Moreover, for every collection of regular parametric submodels
$\{F_{k,t}:t\in(-\epsilon,\epsilon)\}$ through $F_k$, with corresponding scores $s_k\in L_0^2(F_k)$,
$$\left. \frac{d}{dt} \boldsymbol{\rho}(F_{1,t},\ldots,F_{K,t})\right|_{t=0}=\sum_{k=1}^K\E\left\{\boldsymbol h_{\rho,k}(Y_{k})s_k(Y_k)\right\}.$$
\end{assumption}
The lower-bound condition in \ref{assERADE1} guarantees persistent exploration; consistency of the estimated target then follows from the continuity of $\boldsymbol{\rho}$ and the convergence of the treatment-specific empirical CDFs. In analogy with \ref{ass:CLT2}, Condition \ref{assERADE2} is a distributional version of the usual Bahadur-type representation adopted for target-based RAR rules. It does not require the target to depend on a finite-dimensional model parameter, but only that its sequential estimator admits a suitable first-order linear expansion. Condition \ref{assERADE3} requires the influence functions in \eqref{rho-linear} to be the canonical gradients of the functional target and characterizes semiparametric efficiency. The next theorem provides the asymptotic properties of the NP-ERADE and shows that its allocation proportions attain the semiparametric efficiency bound.
\begin{theorem}\label{thmERADE}
Under \ref{assERADE1}, adopting the NP-ERADE in (\ref{NPERADE}), as $n\to\infty$,
\begin{equation}\label{consistERADE}
N_{n,k}\to\infty \qquad\text{and}\qquad \|\widehat F_{n,k}-F_k\|_\infty\to 0 \quad\text{a.s.}  \qquad \text{for }k=1,\ldots,K.
\end{equation}
Thus, $\widehat{\boldsymbol{\rho}}_n\to\boldsymbol{\rho}$ and $\boldsymbol{\pi}_n\to\boldsymbol{\rho}$ a.s. Furthermore, in $\{\ell^\infty(\mathcal X)\}^K$,
\begin{equation}\label{ERADE-FCLT}
\sqrt n \bigl(\widehat F_{n,1}-F_1,\ldots,\widehat F_{n,K}-F_K\bigr)^\top \weak \operatorname{diag}(\boldsymbol{\rho})^{-1/2}\bigl(\mathbb{G}_1\circ F_1,\ldots, \mathbb{G}_K\circ F_K\bigr)^\top
\end{equation}
and if $\Psi$ is Hadamard differentiable at every $F_k$,
\begin{equation}\label{ERADE-score}
\sqrt n(\widehat{\boldsymbol{\Psi}}_n-\boldsymbol{\Psi})\Rightarrow N\left(\boldsymbol 0_K;\boldsymbol V(\boldsymbol{\rho})\right).
\end{equation}
If, in addition, \ref{assERADE2} holds, then
\begin{equation}\label{ERADE-tracking}
\sqrt n(\boldsymbol{\pi}_n-\widehat{\boldsymbol{\rho}}_n)\rightarrow_p \boldsymbol 0_K
\end{equation}
and
\begin{equation}\label{ERADE-target}
\sqrt n (\boldsymbol{\pi}_n-\boldsymbol{\rho})=\sqrt n (\widehat{\boldsymbol{\rho}}_n-\boldsymbol{\rho})+o_p(1) \Rightarrow N(\boldsymbol 0_K;\boldsymbol{\Omega}),
\end{equation}
where $\boldsymbol{\Omega}=\sum_{k=1}^K \rho_k^{-1} \E\left[\boldsymbol h_{\rho,k}(Y_k) \boldsymbol h_{\rho,k}(Y_k)^\top\right]$. If also \ref{assERADE3} holds, $\boldsymbol{\Omega}$ is the semiparametric efficiency bound for estimating $\boldsymbol{\rho}$ and the NP-ERADE allocation proportions attain this bound.
\end{theorem}
\begin{proof}
The proof is given in Appendix \ref{proofTHM3}.
\end{proof}
The ethical target $\boldsymbol{\Lambda}(\boldsymbol{\Psi})$ of Sections \ref{Sec3}-\ref{Sec4} is recovered as a special case. Under \ref{ass:lambda}-\ref{ass:CLT2}, (\ref{eq:strongremainder}) and twice continuous differentiability of $\boldsymbol{\Lambda}$ in a neighborhood of $\boldsymbol{\Psi}$, \ref{assERADE1}-\ref{assERADE2} hold with $\boldsymbol{h}_{\rho,k}(y)=\boldsymbol J\boldsymbol b_k\,\varphi_k(y)$, where $\boldsymbol b_1,\ldots,\boldsymbol b_K$ is the canonical basis of $\mathbb R^K$. Consequently,
$$\boldsymbol{\Omega}=\boldsymbol J \operatorname{diag}\left(\frac{\E\{\varphi_k(Y_k)^2\}}{\Lambda_k(\boldsymbol{\Psi})}\right)_{k=1,\ldots,K}\boldsymbol J^\top=
\boldsymbol J\boldsymbol{\Sigma}\boldsymbol J^\top.$$
If, in addition, $\varphi_k$ is the canonical gradient of $\Psi(F_k)$ in the unrestricted nonparametric model for every $k$ (as for the score functionals in Examples \ref{ex_tau}-\ref{ex_q}), then Condition \ref{assERADE3} is also satisfied and $\boldsymbol J\boldsymbol{\Sigma}\boldsymbol J^\top$ is the semiparametric efficiency bound for $\boldsymbol{\Lambda}(\boldsymbol{\Psi})$.

More generally, the same NP-ERADE mechanism can be employed to track the inferentially optimal targets introduced in Section \ref{Sec5} and, whenever the target admits a plug-in estimator satisfying \ref{assERADE1}-\ref{assERADE2}-\ref{assERADE3}, Theorem \ref{thmERADE} applies directly. For some distributional targets on unbounded supports, however, continuity with respect to the product sup-norm may fail (as, for instance, the functional $A$-optimal target $\boldsymbol{\tilde{\rho}}^A$ in (\ref{funcAottalpha}) depending on $\mathcal{A}_k$s). Moreover, targets involving additional distributional quantities (such as the $A$- and $E$-optimal designs $\boldsymbol\rho^A$ and $\boldsymbol\rho^E$ in (\ref{Aottalpha})) require suitable consistent estimators. In such cases, \ref{assERADE1} can be replaced by the more direct requirement of strong consistency of the target estimator:
\begin{assumptionbis}\label{assc5bis}
The estimator $\widehat{\boldsymbol\rho}_n$ is strongly consistent, namely $\widehat{\boldsymbol\rho}_n\rightarrow \boldsymbol\rho$ a.s.
\end{assumptionbis}
Since $\boldsymbol\rho\in\mathcal S_K$, C4' also guarantees the persistent-exploration of all arms; moreover, it can be established directly through
target-specific arguments and may hold even when the target is not continuous with respect to the product sup-norm or for targets involving pointwise density evaluations
that admit consistent nonparametric estimators (once infinite sampling of every arm has been established, by the independence-after-adaptive-allocation property, standard consistency results for kernel density estimators can be applied to each treatment-specific subsequence). If C4' holds, conclusions in (\ref{consistERADE}) remain valid and therefore $\boldsymbol\pi_n\to\boldsymbol\rho$ a.s. As a consequence, limit (\ref{ERADE-FCLT}) still holds, as does (\ref{ERADE-score}) whenever $\Psi$ is Hadamard differentiable. By contrast, the asymptotic results in (\ref{ERADE-tracking}) and  (\ref{ERADE-target}) require the verification of \ref{assERADE2}, whereas the semiparametric efficiency conclusion additionally requires \ref{assERADE3}. For targets involving pointwise density evaluations, however, \ref{assERADE2}-\ref{assERADE3} generally need not hold in the unrestricted nonparametric model, since density evaluation is not a root-$n$ regular pathwise differentiable functional; moreover, consistently estimating such targets by kernel smoothing of the densities typically involves a slower nonparametric rate.

\begin{example}\label{remERADE}
Assuming the functional $A$-optimal target in (\ref{funcAottalpha}), $\mathcal{A}_k=\frac12 \E|Y_k-Y_k'|$, where $Y_k'$ is an independent copy of $Y_k$. When the outcome space is unbounded, $\boldsymbol{\tilde{\rho}}^A$ need not be continuous with respect to the product sup-norm, since uniform convergence of the treatment CDFs does not control probability mass escaping into the tails. Nevertheless, Conditions C4', \ref{assERADE2} and \ref{assERADE3} still hold if $\mathcal{A}_k>0$ and $\E|Y_k|^{2+\epsilon}<\infty$ for some $\epsilon>0$ and the initialization satisfies
$\widehat{\mathcal{A}}_{n_0,k}>0$ a.s. Indeed, the empirical plug-in estimator of each $\mathcal{A}_k$ is a $V$-statistic:
$$\widehat{\mathcal{A}}_{n,k}=\frac{1}{2N_{n,k}^2}\sum_{i=1}^n\sum_{j=1}^n\delta_{ik}\delta_{jk}|Y_{ik}-Y_{jk}|.$$
To verify C4', we first show that every treatment is sampled infinitely often; suppose by contradiction that $N_{n,k}$ remains bounded for some $k$. Provided that the initialization yields $\widehat{\mathcal A}_{n_0,k}>0$ for every $k$, the corresponding $\widehat{\mathcal A}_{n,k}$ remains equal to a strictly positive constant for every arm sampled only finitely often, whereas, for each treatment sampled infinitely often, the strong law for $V$-statistics yields $\widehat{\mathcal A}_{n,j}\to\mathcal A_j<\infty$ a.s. Consequently, $\liminf_n\widehat\rho_{n,k}>0$ for every finitely sampled arm. Since, from (\ref{NPERADE}), $\Pr(\delta_{n+1,k}=1\mid\mathcal F_n)\geq\gamma\widehat\rho_{n,k}$, the conditional Borel-Cantelli lemma gives $N_{n,k}\to\infty$ a.s., a contradiction. Hence $N_{n,k}\to\infty$ a.s. for every $k=1,\ldots,K$. Thus, by combining the independence-after-adaptive-allocation property of Section \ref{Sec2} with the strong law for $V$-statistics, $\widehat{\mathcal{A}}_{n,k} \rightarrow \mathcal{A}_k$ a.s. for every $k$ and therefore the corresponding plug-in estimator of the $A$-optimal target is also strongly consistent since the functional $(\mathcal{A}_1,\ldots,\mathcal{A}_K)\rightarrow \boldsymbol{\tilde{\rho}}^A$ is continuous. Moreover, the asymptotic linear representation in \ref{assERADE2} follows from the Hoeffding decomposition of the corresponding $V$-statistic and the delta method applied to the map $(\mathcal{A}_1,\ldots,\mathcal{A}_K)\rightarrow \boldsymbol{\tilde{\rho}}^A$. Indeed, the Hoeffding decomposition of $\widehat{\mathcal A}_{n,k}$ has remainder $o_{\mathrm{a.s.}}(N_{n,k}^{-1/2})$. Together with the law of the iterated logarithm for its linear term and $\pi_{n,k}\to\rho_k>0$ a.s., a second-order Taylor expansion of $(\mathcal{A}_1,\ldots,\mathcal{A}_K)\rightarrow \boldsymbol{\tilde{\rho}}^A$ yields $\sqrt n\|\boldsymbol r_n^\rho\|\to 0$ a.s., verifying \ref{assERADE2}.

Moreover, the canonical gradient of $\mathcal{A}_k$ in the unrestricted nonparametric model is $\E|y-Y_k|-2\mathcal{A}_k$ (which is the influence function appearing in the linear term of the Hoeffding decomposition); since
$$\frac{\partial\boldsymbol{\tilde{\rho}}^A}{\partial \mathcal{A}_k}=\frac{\boldsymbol b_k-\boldsymbol{\tilde{\rho}}^A}{2\sqrt{\mathcal{A}_k}(\sum_{j=1}^K\sqrt{\mathcal{A}_j})},$$
the canonical-gradient component of $\boldsymbol{\tilde{\rho}}^A$ associated with perturbations of $F_k$ is
$$\boldsymbol h_{\rho,k}(y)=\frac{\boldsymbol b_k-\boldsymbol{\tilde{\rho}}^A}{2\sqrt{\mathcal{A}_k}(\sum_{j=1}^K\sqrt{\mathcal{A}_j})}\left\{\E|y-Y_k|-2\mathcal{A}_k\right\}.$$
Thus, $\boldsymbol h_{\rho,k}\in \{L_0^2(F_k)\}^K$ is the canonical-gradient component of $\widetilde{\boldsymbol\rho}^{A}$ associated with perturbations of $F_k$, and hence Condition \ref{assERADE3} is satisfied.
\end{example}

\section{Inference at the end of the trial}\label{Sec7}
The asymptotic results of Theorems \ref{thm:tec-clt} and \ref{thmERADE} provide the basis for post-trial inference under both the ethical NP-RAR and the more general NP-ERADE framework. The inferential goal may concern the pre-specified functional scores $\boldsymbol{\Psi}=(\Psi(F_1),\ldots,\Psi(F_K))^\top$ and their contrasts, as well as the comparison of the complete CDFs $F_1,\ldots,F_K$ and their pairwise differences, in order to provide a more detailed description of where and how the treatments differ across the outcome space. For the sake of generality, in what follows we consider the NP-ERADE with a generic target $\boldsymbol{\rho}=\boldsymbol{\rho}(F_1,\ldots,F_K)$, which may represent either an ethical target (as for the ethical NP-RAR, where $\boldsymbol{\rho}=\boldsymbol{\Lambda}(\boldsymbol{\Psi})$) or an inferentially optimal distribution-based target. By Theorem \ref{thm:tec-clt}, all the results of this Section remain valid for the ethical NP-RAR upon replacing $\rho_k$ with $\Lambda_k(\boldsymbol\Psi)$, for $k=1,\ldots,K$.

\subsection{Inference on the pre-specified treatment scores}
For any pair of treatments $k$ and $\ell$, the contrast $\Delta_{k\ell}=\Psi(F_k)-\Psi(F_\ell)$ is naturally estimated by $\widehat\Delta_{n,k\ell}=\Psi(\widehat F_{n,k})-\Psi(\widehat F_{n,\ell})$ and, from (\ref{ERADE-score}),
$\sqrt n (\widehat\Delta_{n,k\ell}-\Delta_{k\ell})\Rightarrow N(0,\rho_k^{-1}\nu_k^2 +\rho_\ell^{-1}\nu_\ell^2)$.
For each $k$, let $\widehat\nu_{n,k}^2$ be a consistent estimator of $\nu_k^2$, so the Wald-type statistic
\begin{equation}\label{wald}
T_{n,k\ell}=\frac{ \widehat\Delta_{n,k\ell}-\Delta_{k\ell} }{\sqrt{\frac{\widehat\nu_{n,k}^2}{N_{n,k}}+\frac{\widehat\nu_{n,\ell}^2}{N_{n,\ell}}}}\Rightarrow N(0;1)
\end{equation}
could be naturally employed for confidence intervals and testing hypotheses of the form $H_0:\Delta_{k\ell}\leq \Delta_0$ vs $H_1:\Delta_{k\ell}> \Delta_0$ for a pre-specified constant $\Delta_0 \in \mathds{R}$. Clearly, the interpretation of $\Delta_{k\ell}$ as well as the choice of the estimator $\widehat\nu_{n,k}^2$ depends on the chosen score. For instance, assuming $\Psi^\tau$, $\Delta_{k\ell}=\Pr(Y_k>\tau)-\Pr(Y_\ell>\tau)=F_\ell(\tau)-F_k(\tau)$ is the difference between the CDFs at $\tau$; moreover, since $\nu_k^2=F_k(\tau)[1-F_k(\tau)]$, then $\widehat\nu_{n,k}^2=\widehat{F}_{n,k}(\tau)[1-\widehat{F}_{n,k}(\tau)]$. For the quantile score $\Psi^\alpha$, $\Delta_{k\ell}=F_k^{-1}(\alpha)-F_\ell^{-1}(\alpha)$ and under the regularity conditions of Example \ref{ex_q} $\nu_k^2=\alpha(1-\alpha)/[f_k\{\Psi^\alpha(F_k)\}]^2$, which could be estimated by
\begin{equation}\label{kernel}
\widehat{\nu}_{n,k}^2=\frac{\alpha(1-\alpha)}{\left[\widehat f_{n,k}\left\{\widehat F_{n,k}^{-1}(\alpha)\right\}\right]^2},
\end{equation}
where $\widehat f_{n,k}$ is a consistent kernel estimator of the density in the $k$th arm. For $\Psi^L$ in Example \ref{ex_L}, $\Delta_{k\ell}=\E(Y_k)-\E(Y_\ell)$ and $\nu_k^2=\var(Y_k)$ could be consistently estimated via the corresponding treatment-specific sample variance.

For $K>2$ treatments, inference can be formulated in terms of pre-specified linear contrasts of the treatment scores: for a fixed $(q\times K)$ contrast matrix $\boldsymbol C$, from Theorems \ref{thm:tec-clt} and \ref{thmERADE}, $\sqrt n \boldsymbol C(\widehat{\boldsymbol{\Psi}}_n-\boldsymbol\Psi)\Rightarrow N\left(\boldsymbol 0_q,\boldsymbol C\boldsymbol V(\boldsymbol\rho) \boldsymbol C^\top\right)$. This formulation covers, for example, comparisons of all experimental treatments with a common control, selected pairwise contrasts, or all pairwise treatment comparisons. When several contrasts are considered simultaneously, multiplicity should be accounted for. The joint Gaussian limit above provides a natural basis for simultaneous confidence intervals or max-type testing procedures, particularly useful to identify treatments whose score is not statistically distinguishable from the largest score, without reducing the analysis to a single estimated ranking.

\subsection{Distributional inference}\label{Sec7.2}
A distinctive feature of the proposed framework is that inference need not be restricted to the scores but may involve the complete CDFs. For two treatments $k$ and $\ell$, the distributional contrast is $D_{k\ell}(x)=F_k(x)-F_\ell(x)$ for $x\in\mathcal X$, with empirical counterpart $\widehat D_{n,k\ell}(x)=\widehat F_{n,k}(x)-\widehat F_{n,\ell}(x)$. From Theorem \ref{thmERADE}, $\sqrt n\{\widehat D_{n,k\ell}-D_{k\ell}\}\Rightarrow W_{k\ell}$ in $\ell^\infty(\mathcal X)$, where $W_{k\ell}(x)=\rho_k^{-1/2}\mathbb G_k\left(F_k(x)\right)-\rho_\ell^{-1/2}\mathbb G_\ell\left(F_\ell(x)\right)$; thus, for each fixed $x$,
$$\var(W_{k\ell}(x))=\frac{F_k(x)\{1-F_k(x)\}}{\rho_k}+\frac{F_\ell(x)\{1-F_\ell(x)\}}{\rho_\ell}.$$
Therefore, pointwise confidence intervals for distributional treatment differences follow directly, whereas simultaneous confidence bands may be constructed from the limiting Gaussian process, providing information that cannot in general be recovered from a single score. In particular, they show whether a treatment advantage is concentrated in a specific region of $\X$ or persists throughout $\X$, especially relevant when the empirical CDFs cross. This also suggests secondary analyses based on stochastic ordering. Since larger outcomes are preferable, treatment $k$ first-order stochastically dominates treatment $\ell$ if $D_{k\ell}(x)=F_k(x)-F_\ell(x)\leq 0$ for every $x\in\mathcal X$, with strict inequality for at least some values of $x$. The estimated difference $\widehat D_{n,k\ell}$, together with a simultaneous confidence band, can therefore provide a useful descriptive assessment of whether the observed distributional evidence is compatible with such a global ordering. In particular, a one-sided band that does not exceed zero over a clinically relevant region of $\X$, while being strictly below zero somewhere, provides evidence of a distributional advantage of treatment $k$ over that region. Conversely, systematic sign changes in the estimated distributional contrast may suggest that treatment superiority is criterion-dependent, although formal testing of stochastic dominance requires dedicated procedures.

When the treatment distributions cross, the crossing locations may themselves be clinically meaningful. Suppose, for simplicity, that $D_{k\ell}$ has an isolated crossing point $x^\star$ such that $D_{k\ell}(x^\star)=0$ and that $F_k$ and $F_\ell$ admit densities continuous in a neighborhood of $x^\star$, with $f_k(x^\star)\neq f_\ell(x^\star)$. Let $\widetilde D_{n,k\ell}$ denote a suitable local interpolation or smoothing of $\widehat D_{n,k\ell}$, and let $\widehat x_n^\star$ be a consistent local zero, i.e.
$\widetilde D_{n,k\ell}(\widehat x_n^\star)=0$. Provided that the chosen interpolation/smoothing procedure yields a consistent local root and is first-order equivalent to the empirical CDF contrast in a neighborhood of $x^\star$, the usual implicit-function expansion yields
\begin{equation}\label{cross1}
\sqrt n(\widehat x_n^\star-x^\star)=-\frac{\sqrt n\{\widehat D_{n,k\ell}(x^\star)-D_{k\ell}(x^\star)\}}{f_k(x^\star)-f_\ell(x^\star)}+o_p(1),
\end{equation}
with asymptotic variance given by
\begin{equation}\label{cross2}
\left(\dfrac{F_k(x^\star)\{1-F_k(x^\star)\}}{\rho_k}+\dfrac{F_\ell(x^\star)\{1-F_\ell(x^\star)\}}{\rho_\ell}\right)\left\{f_k(x^\star)-f_\ell(x^\star)\right\}^{-2}.
\end{equation}
A systematic comparison of alternative interpolation and smoothing procedures constitutes a natural direction for future research.

\section{Finite sample properties}\label{Sec8}
In this section we investigate the finite-sample behavior of the proposed procedures through a simulation study designed to address four complementary aspects of the proposed methodology. First, we assess the finite-sample cost of estimating treatment distributions nonparametrically when a correctly specified parametric model is available. Second, we investigate robustness to model misspecification, with particular emphasis on its consequences for both adaptive allocation and post-trial inference. Third, we consider a multi-arm setting in which the treatment responses belong to different distributional families and clinically relevant differences cannot be captured by their means. Finally, we compare ethical and inferential distribution-based targets when the inferential object is the collection of complete treatment distributions, thereby illustrating how different trial objectives may induce substantially different allocation targets.

Whenever an ethical score is considered, the corresponding target is $\boldsymbol\rho=\boldsymbol\Lambda(\boldsymbol\Psi)$. Depending on the scenario, we compare the ethical NP-RAR of Section \ref{Sec3}, the NP-ERADE of Section \ref{Sec6} targeting either ethical or inferentially optimal distribution-based allocations, and parametric ERADE (P-ERADE) benchmarks \cite{Hu09,Hu2025}. For the P-ERADE in Scenarios I-III, the same parametric working family is assumed across treatment arms, with treatment-specific parameters estimated sequentially and the pre-specified score evaluated on the resulting fitted distributions. Thus, whenever the P-ERADE and the NP-ERADE target the same allocation, they differ only in the way the treatment distributions, and hence the target, are estimated, providing a direct assessment of the effect of parametric modeling assumptions on RAR.

All procedures start from a balanced initialization phase with $10$ observations per treatment. We consider $10,000$ independent Monte Carlo replications for sample sizes $n=100$, $250$ and $500$. The parameter $\gamma$ in (\ref{NPERADE}) is kept fixed across scenarios, with $\gamma=0.5$, whereas $\eta$ in (\ref{allfuncK}) is calibrated separately in each ethical scenario according to the scale of the corresponding score and the desired ethical skew, with $\eta\in\{1,2,5\}$. Allocation behavior is summarized through the empirical mean of the allocation proportions and their empirical standard deviation (SD) or root mean square error (RMSE). For multi-arm settings we use $\operatorname{RMSE}(\boldsymbol\pi_n)=[\E\left\{\|\boldsymbol\pi_n-\boldsymbol\rho\|^2\right\}]^{1/2}$, where $\boldsymbol\rho$ denotes the target under the true treatment distributions. When the inferential object is a pre-specified score vector, estimation accuracy is measured through $\operatorname{RMSE}(\boldsymbol{\widehat\Psi}_n)=[\E\{\|\boldsymbol{\widehat\Psi}_n-\boldsymbol\Psi\|^2\}]^{1/2}$ or the RMSE of relevant pairwise contrasts. For Scenarios I–III, post-trial inference primarily concerns Wald-type procedures for relevant pairwise score contrasts developed in Section \ref{Sec7}; Scenario III additionally illustrates distributional inference on a crossing location as discussed in Section \ref{Sec7.2}. In particular, we report the empirical coverage probabilities (CP) of the two-sided $95\%$ Wald confidence intervals, together with the empirical rejection probability of the corresponding one-sided tests at nominal level $0.05$ (i.e., the empirical type-I error and power). In Scenario IV, where the inferential goal regards the collection of treatment CDFs, accuracy is instead assessed through the total integrated squared error of their empirical estimators
$$\operatorname{TISE}_{n}=\sum_{k=1}^K\int_{\mathbb R}\left\{\widehat F_{n,k}(x)-F_k(x)\right\}^2dx.$$
Gaussian kernel density estimators are used for quantile or crossing-point inference, with normal-reference bandwidths. Additional simulation results related to type-I error under Scenarios I-III can be found at the end of this section.  

\subsection{Scenario I: correctly specified parametric model}
The first scenario investigates whether a substantial finite-sample price is paid for avoiding parametric assumptions when the working model is correctly specified. We reconsider the Gaussian setting of Example \ref{ex2.1}, namely $Y_1\sim N(0.5;2.25)$ and $Y_2\sim N(0.3;0.25)$, under the score $\Psi^\alpha(F_k)=F_k^{-1}(\alpha)$ with $\alpha=0.2$. As shown previously, $\Psi^{0.2}(F_1)= -0.762$ and $\Psi^{0.2}(F_2)= -0.121$, so that $\Delta_{21}= 0.641$; thus, although treatment 1 has the larger mean, treatment 2 is preferable according to $\Psi^\alpha$ and assuming the ethical target with $\eta=1$ we get $\boldsymbol\Lambda(\boldsymbol\Psi^\alpha)=(0.345,\;0.655)^\top$.

We compare the ethical NP-RAR, the ethical NP-ERADE and a parametric ERADE (P-ERADE) based on a Gaussian working model. For the nonparametric procedures,
the treatment scores are estimated sequentially by the corresponding empirical quantiles. For the parametric ERADE, at each step the treatment-specific Gaussian parameters $(\mu_{k},\sigma^2_{k})$ are estimated through the MLEs $(\widehat\mu_{n,k},\widehat\sigma^2_{n,k})$, so the parametric estimate of the score is $\widehat\mu_{n,k}+z_\alpha\widehat\sigma_{n,k}$, where $z_\alpha$ denotes the $\alpha$-quantile of the standard normal distribution. Since the working model is correctly specified, both the parametric estimator and the empirical quantile consistently estimate the same score. Therefore, the P-ERADE, the NP-ERADE and the ethical NP-RAR converge to the same target, providing a benchmark for the finite-sample cost of nonparametric estimation. The designs are first analyzed nonparametrically through the empirical quantile contrast $\widehat\Delta_{n,21}$. For Wald-type inference, the asymptotic variance of the treatment-specific sample quantile is estimated through
$\widehat\nu_{n,k}^2$ in (\ref{kernel}), where $\widehat f_{n,k}$ is a Gaussian kernel density estimator; besides CP of Wald confidence intervals, we report the power of the one-sided test $H_0:\Delta_{21}\leq 0$ versus $H_1:\Delta_{21}>0$. For the P-ERADE, we additionally consider the fully parametric estimator $\widehat\Delta^P_{n,21}=\left(\widehat\mu_{n,2}+z_\alpha\widehat\sigma_{n,2}\right) -\left(\widehat\mu_{n,1}+z_\alpha\widehat\sigma_{n,1}\right)$, to distinguish the gain due to the parametric allocation mechanism from the additional gain obtained by exploiting the correctly specified model also at the analysis stage.

\begin{table}[H]
\centering
\caption{Finite-sample performance under a correctly specified Gaussian model assuming $\Psi^{\alpha}$ and the ethical target. The values of RMSE, CP and Power refer to the nonparametric analysis (the values in parentheses refer to the parametric analysis for the P-ERADE).}
\label{tab:scenario1}
\begin{tabular}{ccccccc}
\toprule
$n$ & Procedure
& $\E(\pi_{n,2})$
& $\mathrm{SD}(\pi_{n,2})$
& RMSE$(\widehat\Delta_{n,21})$
& CP
& Power \\
\midrule

100
& NP-RAR
& 0.631 & 0.084 & 0.416 & 0.95 & 0.567 \\
& NP-ERADE
& 0.666 & 0.085 & 0.480 & 0.95 & 0.570 \\
& P-ERADE
& 0.648 & 0.068 & 0.407 & 0.94 & 0.547 \\
&
&
&
& (0.324) & (0.94) & (0.740) \\
\addlinespace
\midrule
250
& NP-RAR
& 0.651 & 0.067 & 0.262 & 0.95 & 0.913 \\
& NP-ERADE
& 0.662 & 0.058 & 0.278 & 0.95 & 0.920 \\
& P-ERADE
& 0.653 & 0.044 & 0.248 & 0.95 & 0.905 \\
&
&
&
& (0.201) & (0.95) & (0.984) \\
\addlinespace
\midrule
500
& NP-RAR
& 0.655 & 0.052 & 0.177 & 0.95 & 0.995 \\
& NP-ERADE
& 0.658 & 0.039 & 0.179 & 0.95 & 0.997 \\
& P-ERADE
& 0.654 & 0.031 & 0.171 & 0.95 & 0.996 \\
&
&
&
& (0.139) & (0.95) & (1.000) \\
\bottomrule
\end{tabular}
\end{table}
Table \ref{tab:scenario1} shows that all three RAR procedures accurately approach the same ethical target, even for small samples. As expected under correct model specification, the P-ERADE exhibits the smallest allocation variability. The NP-ERADE also exhibits lower allocation variability than the ethical NP-RAR for the larger sample sizes, consistent with the asymptotic results in Theorems \ref{thm:tec-clt} and \ref{thmERADE}. When the final analysis is performed nonparametrically, however, the three procedures have essentially the same inferential precision. Indeed, since they share the same limiting target, the smaller allocation variability of the ERADE procedures does not translate into a first-order reduction of the asymptotic variance of the empirical-quantile contrast. The finite-sample results are consistent with this observation: the RMSEs become increasingly similar as $n$ grows, while the NP-ERADE has power slightly larger than the P-ERADE at the considered sample sizes. Moreover, the kernel-based estimator of the density at the empirical quantile performs well throughout: the empirical coverage probabilities remain close to the nominal $95\%$ level for all procedures and sample sizes.

As regards the P-ERADE, the gain from exploiting the correctly specified Gaussian model is substantial. For instance, at $n=250$ the RMSE decreases from $0.248$ for the
nonparametric analysis to $0.201$ for the fully parametric analysis, while the corresponding power increases from $0.905$ to $0.984$. Hence, this scenario clearly separates two distinct effects. Correct parametric modeling improves the stability with which the allocation target is sequentially approached; however, when nonparametric inference is used at the end of the trial, a reduction in the variability of the allocation proportions does not, by itself, improve the precision of final analysis when the limiting target is unchanged. A further efficiency gain arises only when the correctly specified parametric model is also exploited for post-trial inference.

\subsection{Scenario II: parametric model misspecification}
Assuming the ethical target $\boldsymbol\Lambda(\boldsymbol\Psi)$ with $\eta=5$, the second scenario is designed to isolate the effect of model misspecification while keeping the criterion fixed across procedures. We consider $Y_1\sim N(1;0.16)$ and $Y_2\sim 0.9N(1.5;0.04)+0.1N(-3.5;0.04)$, assuming the linear score $\Psi^L(F_k)=\int_{0}^{1}\left[1-F_k(x)\right]dx$. The treatment distributions have equal expectation $\E(Y_1)=\E(Y_2)=1$, but differ substantially in shape. Under the true treatment CDFs, $\Psi^L(F_1)= 0.841$ and $\Psi^L(F_2)= 0.9$, so treatment 2 is preferable. For the P-ERADE, a Gaussian working model is fitted to both treatment arms and the score $\Psi^L$ is evaluated on the fitted distributions. The working model is correctly specified for treatment 1 but misspecified for treatment 2. In particular, the fitted distribution for treatment 2 converges to $\widetilde F_2=N(1;2.29)$ and $\Psi^L(\widetilde F_2)= 0.627<\Psi^L(F_1)$. Thus, the P-ERADE converges to the misspecified target $\boldsymbol{\Lambda}(\Psi^L(F_1),\Psi^L(\widetilde F_2))=(0.745,0.255)$ favoring treatment 1, whereas the NP-ERADE asymptotically approaches the true target $\boldsymbol{\rho}=\boldsymbol{\Lambda}(\Psi^L(F_1),\Psi^L(F_2)) =(0.427,0.573)$, which is based on the true treatment distributions and favors treatment 2.

The comparison between the ethical NP-RAR and the NP-ERADE assesses the effect of the different allocation mechanisms when the target is correctly estimated, whereas the comparison with the parametric ERADE assesses the effect of model misspecification. We report in Table \ref{tab:scenario2} the same operating characteristics considered in Scenario I (replacing $\mathrm{SD}(\pi_{n,2})$ by $\mathrm{RMSE}(\pi_{n,2})$ evaluated with respect to the true ethical target, in order to emphasize the anomalous behavior of the P-ERADE); as previously, for the P-ERADE, we additionally report model-based inference to evaluate the combined effect of using the misspecified working model for both allocation and final analysis.

\begin{table}[h]
\centering
\caption{Finite-sample performance under a misspecified Gaussian model assuming the score $\Psi^L$ and the ethical target. Allocation RMSE is computed with respect to the true ethical target. For P-ERADE, values in parentheses correspond to Gaussian model-based inference.}
\label{tab:scenario2}
\small
\begin{tabular}{lcccccc}
\toprule
$n$ & Procedure & $\E(\pi_{n,2})$ & $\mathrm{RMSE}(\pi_{n,2})$ & $\mathrm{RMSE}(\widehat\Delta_{n,21})$ & CP & Power \\
\midrule
100 & NP-RAR & 0.555 & 0.082 & 0.057 & 0.96 & 0.314 \\
 & NP-ERADE & 0.568 & 0.070 & 0.059 & 0.95 & 0.322 \\
 & P-ERADE & 0.291 & 0.300 & 0.077 & 0.96 & 0.268   \\
  &   &   &    & (0.301) & (0.06) & (0.003) \\
% & CR & 0.500 & 0.085 & 0.054 & 0.943 & 0.319 & -- & -- & -- \\
\addlinespace
\midrule
250 & NP-RAR & 0.565 & 0.060 & 0.035 & 0.95 & 0.539 \\
 & NP-ERADE & 0.571 & 0.042 & 0.035 & 0.95 & 0.551 \\
 & P-ERADE & 0.269 & 0.310 & 0.047 & 0.96 & 0.410 \\
  &   &  &   & (0.286) & (0.00) & (0.000) \\
% & CR & 0.500 & 0.079 & 0.034 & 0.948 & 0.542 & -- & -- & -- \\
\addlinespace
\midrule
500 & NP-RAR & 0.569 & 0.045 & 0.024 & 0.95 & 0.783 \\
 & NP-ERADE & 0.572 & 0.029 & 0.024 & 0.95 & 0.792 \\
 & P-ERADE & 0.263 & 0.313 & 0.031 & 0.95 & 0.597 \\
  &   &   &   &  (0.279) & (0.00) & (0.000) \\
% & CR & 0.500 & 0.076 & 0.024 & 0.951 & 0.781 & -- & -- & -- \\
\bottomrule
\end{tabular}
\end{table}
The results highlight the distinct behavior of the three RAR procedures under model misspecification. Both the ethical NP-RAR and the NP-ERADE approach the true target, with the NP-ERADE displaying more accurate target tracking: its allocation RMSE decreases from $0.070$ at $n=100$ to $0.029$ at $n=500$, compared with $0.082$ and $0.045$, respectively, for the ethical NP-RAR. The P-ERADE behaves differently because the Gaussian working model reverses the true treatment ranking. The average allocation to treatment 2 approaches the misspecified target rather than the true one. Accordingly, the allocation RMSE remains large as $n$ increases. When the final analysis is performed nonparametrically, allocation toward the misspecified target results in a loss of precision and power. At $n=500$, for example, the empirical power under the P-ERADE is $0.597$, compared with $0.792$ under the NP-ERADE.

Hence, this scenario separates two distinct consequences of model misspecification. When nonparametric inference is used at the end of the trial, misspecification affects the final comparison indirectly through allocation toward an incorrect target, resulting in a loss of precision and power. In practice, however, if the Gaussian model were regarded as sufficiently reliable to guide treatment allocation, it would also be natural to use the same model for the final analysis. In that case, an additional and more severe effect arises: the parametric estimator targets a different contrast from the true $\Delta_{21}$, leading to persistent estimation error and a breakdown of inferential validity. In particular, the coverage of the nominal $95\%$ confidence interval is $0.06$ at $n=100$ and essentially vanishes for larger sample sizes. Thus, in contrast to Scenario I, where correct parametric specification yields an efficiency gain, under model misspecification the use of the parametric model at both the allocation and analysis stages can substantially impair the final treatment comparison.

Notice that, because the working model of a parametric RAR is pre-specified and drives allocation from the earliest stages of the trial, a data-driven modification of that model during the experiment would alter the allocation rule and would generally require a corresponding reassessment of its inferential properties.

\subsection{Scenario III: heterogeneous treatment distributions}
The third scenario considers a multi-arm experiment with three treatments having response distributions from different families. Specifically, let $Y_1\sim 0.95U(-0.2,0)+0.05U(1.8,2)$, $Y_2\sim \operatorname{Exp}(1)-1$ and $Y_3\sim U(-1,1)$, so that all distributions have mean zero. We consider the threshold score $\Psi^\tau(F_k)=\Pr(Y_k>\tau)$ with $\tau=0$. The corresponding true scores are $\Psi^\tau(F_1)=0.05$, $\Psi^\tau(F_2)=e^{-1}= 0.368$ and $\Psi^\tau(F_3)=0.5$, so that treatment 3 is the best and treatment 1 the worst according to the pre-specified criterion. Using the ethical allocation map $\boldsymbol\Lambda$ in (\ref{allfuncK}) with $\eta=2$, the corresponding ethical target is $\boldsymbol\Lambda(\boldsymbol\Psi^\tau)=(0.187,0.353,0.460)^\top$.

As a parametric benchmark, we consider a P-ERADE based on treatment-specific Gaussian working models. Since the three responses have mean zero, the fitted Gaussian models converge respectively to three centered normal distributions and therefore the P-ERADE converges to balance instead of the ethical target (due to the symmetry, the three fitted distributions have the same score $0.5$). Thus, the P-ERADE is asymptotically unable to distinguish the treatments according to the chosen criterion and the corresponding $\operatorname{RMSE}(\boldsymbol\pi_n)$ directly quantifies the allocation distortion induced by model misspecification. We also consider the $A$-optimal target $\boldsymbol\rho^A=(0.182,0.402,0.417)^\top$ for the estimation of the score vector $\boldsymbol\Psi^\tau$ in (\ref{Aotttau}). Hence, in this scenario the ethical and inferential objectives are nearly aligned (the two targets are already very close), but they differ from the balanced allocation induced by the misspecified parametric model. We compare the P-ERADE, the NP-RAR, an ethical NP-ERADE targeting $\boldsymbol\Lambda(\boldsymbol\Psi^\tau)$, and an NP-ERADE targeting $\boldsymbol\rho^A$ (denoted by $A$-opt NP-ERADE).

For post-trial inference, we focus on the contrast $\Delta_{32}=0.5-e^{-1}= 0.132$ (sufficiently moderate to provide a non-degenerate assessment of power over the considered sample sizes) and inference is performed nonparametrically. Besides CP of Wald confidence intervals we report the power of the one-sided test $H_0:\Delta_{32}\leq 0$ versus $H_1:\Delta_{32}>0$.

\begin{table}[h]
\centering
\caption{Finite-sample performance under heterogeneous treatment distributions assuming $\Psi^{\tau}$. Post-trial inference is performed through nonparametric analysis.}
\label{tab:scenario3}
\begin{tabular}{ccccccc}
\toprule
$n$ & Procedure & $\E(\boldsymbol\pi_n)$ & $\operatorname{RMSE}(\boldsymbol\pi_n)$ & $\operatorname{RMSE}(\widehat{\boldsymbol{\Psi}}_n)$ & CP & Power \\
\midrule
100 & NP-RAR & $(0.232,0.347,0.421)$ & 0.106 & 0.123 & 0.94 & 0.338 \\
& NP-ERADE & $(0.194,0.350,0.456)$ & 0.071 & 0.122 & 0.93 & 0.341 \\
& P-ERADE & $(0.271,0.358,0.371)$ & 0.172 & 0.127 & 0.93 & 0.328 \\
& $A$-opt NP-ERADE & $(0.203,0.391,0.406)$ & 0.060 & 0.121 & 0.93 & 0.353 \\
\midrule
250 & NP-RAR & $(0.205,0.350,0.445)$ & 0.072 & 0.077 & 0.95 & 0.597 \\
& NP-ERADE & $(0.189,0.353,0.459)$ & 0.043 & 0.077 & 0.95 & 0.605 \\
& P-ERADE & $(0.306,0.345,0.349)$ & 0.183 & 0.080 & 0.95 & 0.545 \\
& A-opt NP-ERADE & $(0.181,0.402,0.417)$ & 0.051 & 0.076  & 0.95 & 0.625 \\
\midrule
500 & NP-RAR & $(0.196,0.351,0.452)$ & 0.053 & 0.054 & 0.95 & 0.848 \\
& NP-ERADE & $(0.188,0.353,0.459)$ & 0.031 & 0.054 & 0.95 & 0.856 \\
& P-ERADE & $(0.327,0.336,0.337)$ & 0.192 & 0.056 & 0.95 & 0.790 \\
& A-opt NP-ERADE & $(0.177,0.404,0.419)$ & 0.045 & 0.054 & 0.95 & 0.860 \\
\bottomrule
\end{tabular}
\end{table}

Table \ref{tab:scenario3} shows a clear effect of model misspecification on allocation behavior. Both NP-ERADE procedures rapidly approach their respective targets, whereas
the NP-RAR converges more slowly to the ethical target. By contrast, the P-ERADE progressively approaches the balanced allocation. Consequently, model misspecification produces a substantial ethical loss: at $n=500$, the proportion of observations assigned to the best treatment is about 12 percentage points lower under the P-ERADE than under the ethical NP-ERADE. This behavior is also clearly reflected in $\operatorname{RMSE}(\boldsymbol\pi_n)$: for the P-ERADE it remains bounded away from zero, while it vanishes for all the other procedures.

Model misspecification also affects inferential precision. The RMSE of the estimated score vector is systematically larger under the P-ERADE than under the ethical NP-ERADE. This agrees with the theoretical comparison of the asymptotic variances: the balanced allocation induced by the Gaussian working model is inefficient for estimating the true threshold-score vector, whereas the ethical target reduces its total asymptotic variance by approximately $8.4\%$.   The results for the contrast $\Delta_{32}$ further highlight the inferential consequences of the allocation mechanism. While CP remains close to the nominal level for all procedures and sample sizes, power is systematically lower under the misspecified P-ERADE. At $n=250$, for instance, power is $0.545$ under the P-ERADE, compared with $0.605$ under the NP-ERADE and $0.625$ under the $A$-opt NP-ERADE (and this gap persists at $n=500$).

As a further illustration of the distributional inference discussed in Section \ref{Sec7.2}, we consider the crossing between the CDFs of treatments 2 and 3. For $x\in(-1,1)$, $F_2(x)=1-\exp\{-(x+1)\}$ and $F_3(x)=(x+1)/2$, so that (besides the boundary point $-1$) the two distributions have a unique interior crossing $x^\star= 0.594$ (with $f_3(x^\star)-f_2(x^\star)\neq 0$). Thus, although treatment 3 is preferable according to the pre-specified score $\Psi^\tau$ at $\tau=0$, it does not stochastically dominate treatment 2 over the whole outcome space. For each replication we estimate the crossing locations from the sign changes of the piecewise-linear interpolation of the empirical contrast: whenever the interpolated contrast changes sign between two consecutive knots, the corresponding zero is taken as an estimated crossing point $\widehat x_n^\star$
(the piecewise-linear interpolation is first-order equivalent to the empirical contrast, since its uniform interpolation error is of order $O_p(N_{n,2}^{-1}+N_{n,3}^{-1})=O_p(n^{-1})=o_p(n^{-1/2})$). Wald confidence intervals are then constructed from (\ref{cross1})-(\ref{cross2}), replacing the CDFs and target allocations by their empirical counterparts and estimating the treatment densities through Gaussian kernel. Table \ref{tab:crossing} reports the percentage of replications in which an interior crossing is identified, the empirical mean and RMSE of $\widehat x_n^\star$ and CP of the nominal $95\%$ confidence interval.

\begin{table}[h]
\centering
\caption{Finite-sample inference for the interior crossing point $x^\star= 0.594$ between the CDFs of treatments 2 and 3 in Scenario III. The crossing rate is the percentage of replications in which the interior crossing is identified; $\E(\widehat x_n^\star)$, RMSE and CP are computed over those replications.}
\label{tab:crossing}
\begin{tabular}{clcccc}
\hline
$n$ & Procedure & Crossing rate & $\E(\widehat x_n^\star)$
& RMSE$(\widehat x_n^\star)$ & CP \\
\hline
100 & NP-RAR         & 94\% & 0.597 & 0.229 & 0.96 \\
    & NP-ERADE       & 94\% & 0.602 & 0.225 & 0.95 \\
    & P-ERADE        & 92\% & 0.594 & 0.231 & 0.94 \\
    & A-opt NP-ERADE & 94\% & 0.595 & 0.225 & 0.96 \\
\hline
250 & NP-RAR         & 99\% & 0.586 & 0.179 & 0.96 \\
    & NP-ERADE       & 99\% & 0.586 & 0.178 & 0.96 \\
    & P-ERADE        & 98\% & 0.583 & 0.186 & 0.96 \\
    & A-opt NP-ERADE & 99\% & 0.584 & 0.174 & 0.97 \\
\hline
500 & NP-RAR         & 100\% & 0.585 & 0.137 & 0.95 \\
    & NP-ERADE       & 100\% & 0.585 & 0.136 & 0.95 \\
    & P-ERADE        & 100\% & 0.581 & 0.147 & 0.96 \\
    & A-opt NP-ERADE & 100\% & 0.583 & 0.134 & 0.96 \\
\hline
\end{tabular}
\end{table}

The crossing is identified with high probability already for moderate sample sizes and essentially always at $n=500$, while the Wald intervals display coverage close to the nominal level. The empirical mean of $\widehat x_n^\star$ remains close to the true crossing throughout, with a bias that is small relative to its sampling variability. The allocation mechanism also affects the precision of crossing-point estimation. In particular, at $n=500$ the RMSE under the P-ERADE is about $10\%$ larger than under the $A$-optimal NP-ERADE, whereas the NP-RAR and the ethical NP-ERADE exhibit RMSEs much closer to that of the $A$-optimal NP-ERADE. Thus, the conclusions obtained for the scores extend to a genuinely distributional feature that did not drive the adaptive allocation, illustrating the broader post-trial inference supported by the empirical-distribution framework. Overall, Scenario III shows that the nonparametric approach may provide a simultaneous ethical and inferential advantage when treatment comparison is determined by genuinely distributional features that cannot be reduced to the finite-dimensional parametric summaries underlying classical RAR procedures.

\subsection{Scenario IV: inferentially optimal distributional targets}
The final scenario discusses the functional optimal targets of Section \ref{Sec5} and illustrates how substantially different targets may arise from the same collection of treatment distributions depending on the objective of the trial. In particular, we contrast the ethical target $\boldsymbol\Lambda(\boldsymbol\Psi)$ with optimal allocations designed for estimation of the complete treatment CDFs and with the classical Neyman allocation for estimation of treatment means. Its purpose is to assess whether the NP-ERADE can accurately estimate and approach these functional optimal targets and whether the resulting allocation improves estimation of those distributions. We consider $K=3$ treatments with $Y_1\sim U(-1,1)$, $Y_2\sim \mathrm{Laplace}(0,1/4)$ and $Y_3\sim t_3$, where $1/4$ denotes the scale parameter of the Laplace distribution and $t_3$ denotes the Student $t$ distribution with 3 degrees of freedom. Thus, $\E(Y_k)=0$ for each $k=1,2,3$, while $\var(Y_1)=1/3$, $\var(Y_2)=1/8$ and $\var(Y_3)=3$. Notice that the regularity requirements discussed in Example \ref{remERADE} are satisfied in the present setting: for each treatment $k$, $\mathcal{A}_k>0$ and $\E|Y_k|^{2+\epsilon}<\infty$ for some $\epsilon>0$ (for the $t_3$ distribution, any $\epsilon\in(0,1)$ can be chosen). Therefore, the empirical functional $A$-optimal target is strongly consistent and its asymptotic linear representation satisfies \ref{assERADE2}.

We first consider the functional optimality criteria for estimation of the complete treatment CDFs. Since all three distributions are continuous, $\mathcal{B}_k=1/4$ for $k=1,2,3$ and therefore the functional $E$-optimal target is balanced. For functional $A$-optimality, recall that $\mathcal{A}_k=\int_{\mathbb R}F_k(x)\{1-F_k(x)\}dx$, in the present setting $\mathcal{A}_1=1/3$, $\mathcal{A}_2=3/16$ and $\mathcal{A}_3=\frac{3\sqrt{3}}{2\pi}$, which yields $\widetilde{\boldsymbol\rho}^A =(0.301,0.226,0.474)^\top$. As a classical benchmark, we also consider the Neyman allocation, which is $A$-optimal for estimation of the treatment means, namely $\boldsymbol\rho^N=(0.217,0.133,0.650)^\top$. In particular, we compare the Neyman ERADE (which sequentially estimates $\boldsymbol\rho^N$ through the treatment-specific sample standard deviations), with the $A$- and $E$-optimal NP-ERADEs (which sequentially estimate $\mathcal{A}_{k}$s and $\mathcal{B}_{k}$s via the empirical CDFs) converging to $\widetilde {\boldsymbol\rho}^A$ and $\widetilde{\boldsymbol\rho}^E$, respectively. Moreover, to illustrate the discrepancy between ethical and inferential objectives, we also consider the ethical NP-RAR under the threshold score $\Psi^{\tau}$ with $\tau=-1$; since $\Pr(Y_1>-1)=1$, $\Pr(Y_2>-1)= 0.991$ and $\Pr(Y_3>-1)= 0.804$, assuming $\eta=5$ the corresponding ethical target is $\boldsymbol\Lambda(\boldsymbol\Psi^\tau)=(0.429,0.410,0.161)^\top$.

\begin{table}[h]
\centering
\caption{Finite-sample performance for estimation of the complete treatment distributions.}
\label{tab:scenario4}
\begin{tabular}{ccccc}
\toprule
$n$ & Procedure & $\E(\boldsymbol\pi_n)$ & $\operatorname{RMSE}(\boldsymbol\pi_n)$ & $\E(\operatorname{TISE}_n)$ \\
\midrule
100 & Neyman ERADE & $(0.234,0.157,0.609)$ & 0.086 & 0.041 \\
& $A$-opt NP-ERADE & $(0.303,0.226,0.471)$ & 0.042 & 0.037 \\
& $E$-opt NP-ERADE & $(0.333,0.333,0.333)$ & 0.014 & 0.040 \\
& NP-RAR & $(0.399,0.385,0.216)$ & 0.109 & 0.056 \\
\midrule
250 & Neyman ERADE & $(0.230,0.136,0.634)$ & 0.060 & 0.017 \\
& $A$-opt NP-ERADE & $(0.302,0.226,0.472)$ & 0.026 & 0.015 \\
& $E$-opt NP-ERADE & $(0.333,0.333,0.333)$ & 0.006 & 0.016 \\
& NP-RAR & $(0.419,0.400,0.181)$ & 0.075 & 0.026 \\
\midrule
500 & Neyman ERADE & $(0.226,0.136,0.638)$ & 0.047 & 0.008 \\
& $A$-opt NP-ERADE & $(0.301,0.226,0.473)$ & 0.018 & 0.007 \\
& $E$-opt NP-ERADE & $(0.333,0.333,0.333)$ & 0.003 & 0.008 \\
& NP-RAR & $(0.425,0.405,0.170)$ & 0.057 & 0.014 \\
\bottomrule
\end{tabular}
\end{table}

Table \ref{tab:scenario4} shows that all considered procedures accurately track their corresponding targets. The $A$-optimal NP-ERADE yields the smallest TISE at every considered sample size, consistent with the criterion used to construct its target, whereas the $E$-optimal NP-ERADE rapidly approaches the balanced allocation and exhibits a slightly larger distribution-estimation error. The Neyman ERADE, instead, progressively approaches an allocation that is optimal for estimation
of the treatment means but not of their complete CDFs, resulting in a larger TISE for the larger sample sizes. A markedly different behavior is observed for the ethical NP-RAR. By favoring treatments 1 and 2 according to the pre-specified ethical criterion, the procedure pays a substantial price in inferential precision, as reflected by its considerably larger TISE, which at $n=500$ is approximately $99\%$ larger than the corresponding value under the $A$-optimal NP-ERADE. Thus, this scenario highlights the trade-off between ethical objectives and accurate estimation of the complete treatment distributions, as well as the dependence of the optimal allocation on the specific inferential criterion.

\subsection{Type-I error under null configurations}
Scenarios I-III assess coverage and power under the alternative hypothesis configurations. We complement these results by investigating the finite sample inferential properties under null configurations.  For a pair of treatments $k$ and $\ell$, inference is based on the Wald statistic $T_{n,k\ell}$ in \eqref{wald}. For Scenarios I-III, type-I error is evaluated at $\Delta_{k\ell}=\Psi(F_k)-\Psi(F_\ell)=0$, considering both an equal-score configuration, in which $\Psi(F_k)=\Psi(F_\ell)$ while $F_k\neq F_\ell$, and the stronger identical-distribution configuration $F_k=F_\ell$. For Scenario I, the equal-score null is constructed by leaving treatment 1 unchanged ($Y_1\sim N(0.5;2.25)$) and modifying only the mean of treatment 2 ($Y_2\sim N(-0.3416;0.25)$) so that $\Psi^{\alpha}(F_1)=\Psi^{\alpha}(F_2)$ for $\alpha=0.2$; while under the stronger null configuration, both treatments are generated from $N(0.5;2.25)$. For Scenario II, treatment 1 is left unchanged and, under the equal-score null, we let $Y_2\sim 0.9N(1.0309;0.04)+0.1N(-3.9691;0.04)$, so that $\Psi^L(F_2)=\Psi^L(F_1)$; while under the identical-distribution null, treatment 2 is generated from $N(1;0.16)$. For Scenario III, type-I error is assessed for $\Delta_{32}=\Psi^\tau(F_3)-\Psi^\tau(F_2)$ with $\tau=0$. The response distributions of treatments 1 and 2 are unchanged, while under the equal-score null we let $Y_3\sim(1-e^{-1})U(-1,0)+e^{-1}U(0,e-1)$, so that $\Pr(Y_3>0)=\Pr(Y_2>0)=e^{-1}$, while $\E(Y_3)=0$. Under the identical-distribution null, $Y_3$ has the same distribution as $Y_2$. Both null configurations therefore share the same score vector and hence the same ethical and $A$-optimal targets. For the P-ERADE, we additionally report the corresponding Gaussian model-based analysis under the null configurations.

The comparison consequently isolates the effect of distributional heterogeneity under the null, while keeping the inferential target unchanged. The results are reported in Tables \ref{tab:typeI_equalscore}-\ref{tab:typeI_identical}.

\begin{table}[h]
\centering
\caption{CP and type-I error under the equal-score null configuration in Scenarios I-III. Unless otherwise specified, inference is based on the nonparametric estimator of the pre-specified score. For the P-ERADE, values in parentheses refer to the corresponding parametric analysis.}
\label{tab:typeI_equalscore}
\begin{tabular}{llcc|cc|cc}
\hline
& & \multicolumn{2}{c|}{Scenario I} & \multicolumn{2}{c|}{Scenario II} & \multicolumn{2}{c}{Scenario III} \\
$n$ & Procedure & CP & Type-I error & CP & Type-I error & CP & Type-I error \\
\hline
100 & NP-RAR    & 0.95    & 0.042     & 0.96    & 0.059   & 0.94   & 0.060 \\
    & NP-ERADE  & 0.95    & 0.047     & 0.96    & 0.056   & 0.94   & 0.058 \\
    & P-ERADE   & 0.95    & 0.040     & 0.96    & 0.060   & 0.93   & 0.061 \\
    &           & (0.94)  & (0.042)   & (0.01)  & (0.001) & (0.9) & (0.097) \\
    & $A$-opt NP-ERADE & -- & --        & --      & --      & 0.94   & 0.058 \\
\hline
250 & NP-RAR   & 0.95     & 0.046     & 0.95    & 0.053   & 0.95   & 0.054 \\
    & NP-ERADE & 0.95     & 0.046     & 0.95    & 0.055   & 0.95   & 0.054 \\
    & P-ERADE  & 0.95     & 0.045     & 0.95    & 0.061   & 0.94   & 0.056 \\
    &          & (0.94)   & (0.047)   & (0.00)  & (0.000) & (0.93) & (0.075) \\
    & $A$-opt NP-ERADE & -- & --        & --      & --      & 0.95   & 0.055 \\
\hline
500 & NP-RAR    & 0.95 & 0.044  & 0.95 & 0.051  & 0.95 & 0.055 \\
    & NP-ERADE  & 0.95 & 0.044 & 0.95 & 0.051 & 0.95 & 0.050 \\
    & P-ERADE   & 0.95 & 0.039 & 0.95 & 0.058  & 0.94 & 0.053 \\
    &           & (0.95) & (0.042)  & (0.00) & (0.000)  & (0.94) & (0.064) \\
    & $A$-opt NP-ERADE & -- & --  & -- & -- & 0.95 & 0.052 \\
\hline
\end{tabular}
\end{table}

\begin{table}[h]
\centering
\caption{CP and type-I error under the identical-distribution null configuration in Scenarios I--III. Unless otherwise specified, inference is based on the nonparametric estimator of the pre-specified score. For the P-ERADE, values in parentheses refer to the corresponding parametric analysis.}
\label{tab:typeI_identical}
\begin{tabular}{llcc|cc|cc}
\hline
& & \multicolumn{2}{c|}{Scenario I}
& \multicolumn{2}{c|}{Scenario II}
& \multicolumn{2}{c}{Scenario III} \\
$n$ & Procedure
& CP & Type-I error
& CP & Type-I error
& CP & Type-I error \\
\hline
100 & NP-RAR & 0.95 & 0.045 & 0.95 & 0.050 & 0.94 & 0.061 \\
    & NP-ERADE & 0.95 & 0.053 & 0.95 & 0.048 & 0.93 & 0.060 \\
    & P-ERADE & 0.96 & 0.046 & 0.95 & 0.052 & 0.93 & 0.062 \\
    & & (0.95) & (0.052) & (0.95) & (0.052) & (0.87) & (0.087) \\
    & $A$-opt NP-ERADE & -- & -- & -- & -- & 0.94 & 0.058 \\
\hline
250 & NP-RAR & 0.95 & 0.048 & 0.95 & 0.048 & 0.94 & 0.056 \\
    & NP-ERADE & 0.95 & 0.046 & 0.95 & 0.047 & 0.95 & 0.051 \\
    & P-ERADE & 0.96 & 0.046 & 0.95 & 0.046 & 0.94 & 0.057 \\
    & & (0.95) & (0.050)  & (0.95) & (0.048) & (0.92) & (0.067) \\
    & $A$-opt NP-ERADE & -- & -- & -- & -- & 0.94 & 0.054 \\
\hline
500 & NP-RAR & 0.95 & 0.048  & 0.95 & 0.048 & 0.94 & 0.054 \\
    & NP-ERADE & 0.95 & 0.053 & 0.95 & 0.045 & 0.95 & 0.050 \\
    & P-ERADE & 0.95 & 0.049 & 0.95 & 0.048 & 0.94 & 0.050 \\
    & & (0.95) & (0.048) & (0.95) & (0.048) & (0.93) & (0.061) \\
    & $A$-opt NP-ERADE & -- & -- & -- & -- & 0.95 & 0.052 \\
\hline
\end{tabular}
\end{table}
Overall, the nonparametric analyses show empirical rejection probabilities generally close to the nominal $0.05$ level and coverage probabilities close to $95\%$. These results support the finite-sample validity of the Wald inference developed in Section \ref{Sec7} under adaptive allocation, both under the stronger identical-distribution null and under the more general equal-score null. The additional Gaussian model-based analyses highlight, however, the consequences of relying on a working parametric model. In Scenario II, under the equal-score null, model misspecification leads to severe undercoverage and a rejection probability converging toward zero as $n$ increases, reflecting the fact that the parametric analysis targets a pseudo-contrast that does not coincide with the true score contrast. In Scenario III, the Gaussian working model yields the same limiting value of the pre-specified score for treatments 2 and 3, so that the corresponding model-based contrast is also equal to zero. Hence, unlike in Scenario II, the parametric estimator remains asymptotically centered at the null, although the corresponding Wald inference shows some finite-sample departure from the nominal level for small samples.

\FloatBarrier
\section{Discussion}\label{Sec9}
This paper provides a general nonparametric framework for distribution-based RAR procedures and post-trial inference. Its practical interpretation depends on specifying the functional that defines treatment performance: when treatment CDFs cross, different trial objectives may legitimately induce different rankings. The allocation target could also reflect an explicit inferential objective and we derive optimal designs for estimating functional scores and complete treatment distributions. Separating target estimation from target tracking clarifies the roles of the two proposed procedures. The ethical NP-RAR uses the estimated target directly as the assignment probability, whereas the NP-ERADE corrects departures of the allocation proportions from that target and, under the stated regularity conditions, attains its semiparametric efficiency bound. This distinguishes allocation variability from the inferential precision achieved by the chosen target. The joint empirical-process theory provides a common basis for design criteria and for inference on distributional features beyond the score used to drive randomization. The framework accommodates multiple treatments and heterogeneous response distributions, with examples involving threshold probabilities, quantiles and integrated distributional criteria. Although motivated by clinical trials, it also applies to sequential online and policy experiments satisfying the same assumptions.

The simulations illustrate the trade-off between efficiency under correct parametric specification and robustness to misspecification.
A misspecified parametric RAR may converge to an incorrect allocation target and may distort the contrast estimated in the final analysis, whereas the nonparametric procedures consistently target the pre-specified distributional criterion across heterogeneous treatment distributions, while retaining valid inference. They also show that optimal targets for score, mean and CDF estimation may differ substantially

Moreover, the proposed methodology opens several directions for future work. First, one can investigate distribution-based targets that explicitly trade off ethical requirements and inferential precision, as well as optimal choices of the score functional $\Psi$ under clinical utility or welfare criteria. Second, one can develop finite-sample and resampling refinements of the post-trial inferential procedures described in Section \ref{Sec7}, particularly for non-smooth functionals and simultaneous distributional comparisons. Third, one can extend the framework to delayed responses, prognostic factors/covariates, survival outcomes and ordinal endpoints. Finally, one can systematically compare and refine interpolation and smoothing procedures for post-trial inference on crossing locations and slopes of distributional contrasts. Moreover, formal testing of stochastic dominance requires dedicated methods for functional inequality constraints, accounting for potentially extended regions where the CDFs coincide.

\appendix
\section{Proofs of the main results}\label{app:proofs}
\subsection{Proof of Theorem \ref{thm:gc}}\label{proofTHM1}
By \ref{ass:lambda}, $\liminf_{n\to\infty}\Lambda_k(\widehat{\boldsymbol{\Psi}}_n)>0$ a.s. for each $k$. Hence, almost surely, there exist $c_k(\omega)>0$ and a random index $n_k(\omega)<\infty$ such that $\Lambda_k(\widehat{\boldsymbol{\Psi}}_n)\ge c_k(\omega)$ for all $n\ge n_k(\omega)$. Thus, standard conditional Borel-Cantelli lemma or martingale arguments imply that $N_{n,k}\to\infty$ a.s. for all $k=1,\ldots,K$. Under adaptive allocation, the subsequence of observed responses assigned to treatment $k$ has the same marginal distributional structure as an i.i.d. sample from $F_k$, by the independence-after-adaptive-allocation argument of B\'elisle and Melfi \cite{belisle_melfi}. Indeed, for every treatment $k$, let $\mathcal{T}_{m,k}=\inf\{n\geq1:N_{n,k}=m\}$ for $m\geq1$ and define $\tilde{Y}_{m,k}=Y_{\mathcal{T}_{m,k},k}$, which denotes the response of the $m$th subject allocated to treatment $k$. Since treatment assignment is based only on the observed past and is independent of the potential outcomes of the current subject, from \cite{belisle_melfi} it follows that $\tilde{Y}_{1,k},\tilde{Y}_{2,k},\ldots$ are i.i.d. with common CDF $F_k$. Moreover, $\widehat F_{n,k}(x)=N_{n,k}^{-1}\sum_{i=1}^{N_{n,k}}\1\{\tilde{Y}_{i,k}\leq x\}$ and since $N_{n,k}\to\infty$ a.s., the Glivenko-Cantelli theorem therefore yields $\|\widehat F_{n,k}-F_k\|_\infty\to 0$ a.s. Consistency of the estimated score vector and of the allocation probabilities follows from continuity of $\Psi$ and $\boldsymbol\Lambda$. Finally, for each $k=1,\dots,K$,
$$\pi_{n,k}=\frac{N_{n_0,k}}{n}+ \frac1n\sum_{i=n_0+1}^n \Lambda_k(\widehat{\boldsymbol{\Psi}}_{i-1})+\frac1n\sum_{i=n_0+1}^n\bigl\{\delta_{ik}-\Lambda_k(\widehat{\boldsymbol{\Psi}}_{i-1})\bigr\}.$$
Due to the finiteness of the initialization phase, almost surely the first term tends to zero, the second one to $\Lambda_k(\boldsymbol{\Psi})$ by Ces\`aro convergence, while the third one tends to zero by the martingale strong law. Therefore, $\boldsymbol\pi_n\to \boldsymbol{\Lambda}(\boldsymbol{\Psi})$ a.s.

\subsection{Proof of Theorem \ref{thm:tec-clt}}\label{proofTHM2}
In what follows, the finite initialization phase is omitted from the summations, since its contribution is asymptotically negligible, and for notational simplicity, all subsequent sums are still written as starting from $1$. For every $k=1,\ldots,K$ and $x\in\mathcal X$, let $\xi_{ik}(x)=\delta_{ik}\left[\1\{Y_{ik}\le x\}-F_k(x)\right]$ and $M_{n,k}(x)=\sum_{i=1}^n \xi_{ik}(x)$. From (\ref{eq:direct-design}), $\E(\xi_{ik}(x)\mid\mathcal F_{i-1})=0$ and therefore, for each fixed $x$, $\{M_{n,k}(x)\}_{n\ge1}$ is a centered martingale with uniformly bounded increments. Thus, standard martingale strong-law arguments yield
\begin{equation}\label{conv}
\widehat F_{n,k}(x)-F_k(x)=\frac{M_{n,k}(x)}{N_{n,k}}\rightarrow 0\quad  a.s.
\end{equation}
for each fixed $x\in\mathcal X$, since $N_{n,k}\to\infty$ a.s. from Theorem \ref{thm:gc}. By using the independence-after-adaptive-allocation structure, the treatment-specific empirical processes inherit the usual Donsker behavior along the treatment-specific subsequences, since $N_{n,k}\rightarrow\infty$ for every $k=1,\ldots,K$ and $\boldsymbol{\pi}_{n}\to \boldsymbol{\Lambda}(\boldsymbol{\Psi})$ a.s. Indeed, at each treatment $k$, for any pair $x,y\in\mathcal X$
$$
\begin{aligned}
\E\{\xi_{ik}(x)\xi_{ik}(y)\mid\mathcal F_{i-1}\}&=\E\Big[\delta_{ik}[\1\{Y_{ik}\le x\}-F_k(x)][\1\{Y_{ik}\le y\}-F_k(y)]\mid\mathcal F_{i-1}\Big]\\
&=\Lambda_k(\widehat{\boldsymbol{\Psi}}_{i-1})\E\Big[[\1\{Y_{ik}\le x\}-F_k(x)][\1\{Y_{ik}\le y\}-F_k(y)]\Big]\\
&=\Lambda_k(\widehat{\boldsymbol{\Psi}}_{i-1})\operatorname{Cov}\left(\1\{Y_{ik}\le x\},\1\{Y_{ik}\le y\}\right)\\
&=\Lambda_k(\widehat{\boldsymbol{\Psi}}_{i-1})\left\{F_k(x\wedge y)-F_k(x)F_k(y)\right\}.
\end{aligned}
$$
By Theorem \ref{thm:gc}, $\Lambda_k(\widehat{\boldsymbol{\Psi}}_n)\rightarrow \Lambda_k(\boldsymbol\Psi)$ a.s., and therefore Ces\`aro convergence yields
$$n^{-1}\sum_{i=1}^n \E\{\xi_{ik}(x)\xi_{ik}(y)\mid\mathcal F_{i-1}\} \rightarrow \Lambda_k(\boldsymbol\Psi) \left\{F_k(x\wedge y)-F_k(x)F_k(y)\right\}\quad a.s.$$
Moreover, for $k\neq\ell$, $\xi_{ik}(x)\xi_{i\ell}(y)=0$, since $\delta_{ik}\delta_{i\ell}=0$, and hence the limiting cross-covariances between different treatment-specific empirical processes are zero. Furthermore, the Lindeberg condition is satisfied because the increments are uniformly bounded. Since the class of indicator functions of half-lines is a standard Donsker class with bounded envelope, the corresponding entropy conditions are satisfied and the empirical-process central limit theorem for $\ell^\infty$-valued martingale difference arrays applies (see, e.g., \citep{Nis2000}), yielding
$$n^{-1/2}\left(M_{n,1},\ldots,M_{n,K}\right)^\top \Rightarrow \operatorname{diag}
\left\{\boldsymbol{\Lambda}(\boldsymbol{\Psi})\right\}^{1/2}\left(\mathbb{G}_1\circ F_1,\ldots,\mathbb{G}_K\circ F_K\right)^\top$$
in $\{\ell^\infty(\mathcal X)\}^K$, where $\mathbb{G}_1,\ldots,\mathbb{G}_K$ are independent standard Brownian bridges, since the joint limit is Gaussian and the limiting cross-covariances between different treatment arms are zero. Since, by Theorem \ref{thm:gc}, $\boldsymbol{\pi}_{n}\to \boldsymbol{\Lambda}(\boldsymbol{\Psi})$ a.s., from (\ref{conv}) and Slutsky's theorem
$$\sqrt n\bigl(\widehat F_{n,k}-F_k\bigr) =\pi_{n,k}^{-1} \left\{\frac{M_{n,k}}{\sqrt{n}}\right\}\Rightarrow
\bigl(\Lambda_k(\boldsymbol\Psi)\bigr)^{-1/2}\mathbb{G}_k\circ F_k$$
jointly for $k=1,\ldots,K$ in $\bigl\{\ell^\infty(\mathcal X)\bigr\}^K$, which proves (\ref{CLTF}).

As regards (\ref{CLTpsi}), if $\Psi$ is Hadamard differentiable at each $F_k$, the functional delta method gives
$$\sqrt n\left\{\Psi(\widehat F_{n,k})-\Psi(F_k)\right\}=\dot\Psi_{F_k}\left\{\sqrt n(\widehat F_{n,k}-F_k)\right\}+o_p(1).$$
Hence, $\sqrt n(\widehat{\boldsymbol{\Psi}}_n-\boldsymbol\Psi)\Rightarrow \boldsymbol Z$ proving (\ref{CLTpsi}). If, in addition, $\boldsymbol\Lambda:\mathbb R^K\to\mathcal S_K$ is differentiable at $\boldsymbol\Psi$,
$$\sqrt n \left\{\boldsymbol\Lambda(\widehat{\boldsymbol{\Psi}}_n)-\boldsymbol\Lambda(\boldsymbol\Psi)\right\}=\sqrt n \boldsymbol J(\widehat{\boldsymbol{\Psi}}_n-\boldsymbol\Psi)+o_p(1)\Rightarrow \boldsymbol J\boldsymbol Z,$$
which proves (\ref{CLTlambda}).

Finally, to prove (\ref{CLTpi}), assume that \ref{ass:CLT2} holds and that $\boldsymbol\Lambda$ is differentiable at $\boldsymbol\Psi$. For every $k$, let $U_{i,k}=\delta_{ik}\varphi_k(Y_{ik})$ and $S_{n,k}=\sum_{i=1}^n U_{i,k}$; thus $\E(U_{i,k}\mid\mathcal F_{i-1})=0$; from \ref{ass:CLT2}, $\E(S_{m,k}^2)=O(m)$ and $\Psi(\widehat F_{n,k})-\Psi(F_k)=S_{n,k}/N_{n,k}+r_{n,k}$. By Theorem \ref{thm:gc}, $\pi_{n,k}\rightarrow \Lambda_k(\boldsymbol\Psi)>0$ a.s. and $\E|S_{m,k}|=O(\sqrt m)$ since $\E(S_{m,k}^2)=O(m)$; so, from the condition on $r_{n,k}$ in \ref{ass:CLT2},
\begin{equation}\label{Op}
\frac1{\sqrt n}\sum_{i=1}^{n}\left|\Psi(\widehat F_{i,k})-\Psi(F_k)\right|=O_p(1)\quad \text{and} \quad \frac1{\sqrt n}\sum_{i=1}^{n}\left\|\widehat{\boldsymbol{\Psi}}_i-\boldsymbol\Psi \right\|=O_p(1).
\end{equation}
Since $\boldsymbol\Lambda$ is differentiable at $\boldsymbol\Psi$ and $\widehat{\boldsymbol{\Psi}}_n\to\boldsymbol\Psi$ a.s., (\ref{Op}) also implies
$$\frac1{\sqrt n}\sum_{i=1}^{n} \left\| \boldsymbol\Lambda(\widehat{\boldsymbol{\Psi}}_i) -\boldsymbol\Lambda(\boldsymbol\Psi)\right\|=O_p(1).$$
Now letting $\boldsymbol e_i=\boldsymbol\delta_i-\boldsymbol\Lambda(\widehat{\boldsymbol{\Psi}}_{i-1})$, then
$\E(\boldsymbol e_i\mid\mathcal F_{i-1})=\boldsymbol0_K$ and $\sum_{i=1}^n\boldsymbol e_i=O_p(\sqrt n)$, since it is a martingale with uniformly bounded increments. Thus,
\begin{equation}\label{decomp}
\boldsymbol{N}_n-n\boldsymbol\Lambda(\boldsymbol\Psi)=\sum_{i=1}^n\boldsymbol e_i+\sum_{i=1}^n\left\{\boldsymbol\Lambda(\widehat{\boldsymbol{\Psi}}_{i-1})-\boldsymbol\Lambda(\boldsymbol\Psi)\right\}+O_p(1)=O_p(\sqrt n).
\end{equation}
Moreover, since $\E|S_{m,k}|=O(\sqrt m)$, it follows that
\begin{equation}\label{p1}
\frac1{\sqrt n}\sum_{m=1}^{n-1}\frac{|S_{m,k}|}{m}=O_p(1).
\end{equation}
By Theorem \ref{thm:gc}, $m/N_{m,k}=\pi_{m,k}^{-1} \rightarrow (\Lambda_k(\boldsymbol\Psi))^{-1}$ a.s. and, from (\ref{p1}),
$$\frac1{\sqrt n}\sum_{m=1}^{n-1}S_{m,k}\left\{\frac1{N_{m,k}}-\frac1{m\Lambda_k(\boldsymbol\Psi)}\right\}=\frac1{\sqrt n}
\sum_{m=1}^{n-1}\frac{S_{m,k}}{m}\left\{\frac{m}{N_{m,k}}-\frac1{\Lambda_k(\boldsymbol\Psi)}\right\}=o_p(1).$$
Consequently,
$$\frac1{\sqrt n}\sum_{i=1}^{n-1}\left\{\Psi(\widehat F_{i,k})-\Psi(F_k)\right\}=\frac1{\sqrt n}\sum_{i=1}^{n-1}\frac{S_{i,k}}{i\Lambda_k(\boldsymbol\Psi)}+o_p(1).$$
Letting now
$$\boldsymbol H_i=\left(\frac{\delta_{i1}\varphi_1(Y_{i1})}{\Lambda_1(\boldsymbol\Psi)},\ldots,\frac{\delta_{iK}\varphi_K(Y_{iK})}{\Lambda_K(\boldsymbol\Psi)}\right)^\top ,$$
then $S_{m,k}/\Lambda_k(\boldsymbol\Psi)=\sum_{i=1}^mH_{i,k}$ and (up to the asymptotically irrelevant initialization of the procedure)
$$\frac1{\sqrt n}\sum_{m=1}^{n-1}(\widehat{\boldsymbol{\Psi}}_m-\boldsymbol\Psi)=\frac1{\sqrt n}\sum_{m=1}^{n-1}\frac1m\sum_{i=1}^m\boldsymbol H_i+o_p(1)=
\frac1{\sqrt n}\sum_{i=1}^{n-1}a_{n,i}\boldsymbol H_i+o_p(1),$$
where $a_{n,i}=\sum_{m=i}^{n-1}m^{-1}$. By differentiability of $\boldsymbol\Lambda$ at $\boldsymbol\Psi$, $\boldsymbol\Lambda(\widehat{\boldsymbol{\Psi}}_m)-\boldsymbol\Lambda(\boldsymbol\Psi)=\boldsymbol J(\widehat{\boldsymbol{\Psi}}_m-\boldsymbol\Psi)+\boldsymbol\zeta_m$,
where $\|\boldsymbol\zeta_m\|=o\left(\|\widehat{\boldsymbol{\Psi}}_m-\boldsymbol\Psi\|\right)$ a.s. Since $\widehat{\boldsymbol{\Psi}}_m\to\boldsymbol\Psi$ a.s., for every $\varepsilon>0$ there exists an a.s. finite random index $m_\varepsilon$ such that $\|\boldsymbol\zeta_m\|\le \varepsilon \|\widehat{\boldsymbol{\Psi}}_m-\boldsymbol\Psi\|$ for $m\ge m_\varepsilon$. Hence,
$$\frac1{\sqrt n}\sum_{m=1}^{n-1}\|\boldsymbol\zeta_m\|\le \frac1{\sqrt n} \sum_{m=1}^{m_\varepsilon-1}\|\boldsymbol\zeta_m\| +\frac{\varepsilon}{\sqrt n}
\sum_{m=m_\varepsilon}^{n-1}\|\widehat{\boldsymbol{\Psi}}_m-\boldsymbol\Psi\|=o_p(1),$$
since the first term converges to zero a.s. and the second one is $\varepsilon O_p(1)$ (therefore $n^{-1/2}\sum_{m=1}^{n-1}\|\boldsymbol\zeta_m\|=o_p(1)$ since $\varepsilon>0$ is arbitrary). Thus,
$$\frac1{\sqrt n}\sum_{m=1}^{n-1}\left\{\boldsymbol\Lambda(\widehat{\boldsymbol{\Psi}}_m)-\boldsymbol\Lambda(\boldsymbol\Psi)\right\}=
\frac1{\sqrt n}\sum_{i=1}^{n-1}a_{n,i}\boldsymbol J\boldsymbol H_i+o_p(1)$$
and therefore, from (\ref{decomp}),
\begin{equation}\label{decomp2}
\sqrt n \left\{\boldsymbol\pi_n-\boldsymbol\Lambda(\boldsymbol\Psi)\right\}=\frac1{\sqrt n}\sum_{i=1}^{n}\boldsymbol e_i+
\frac1{\sqrt n}\sum_{i=1}^{n-1}a_{n,i}\boldsymbol J\boldsymbol H_i+o_p(1),
\end{equation}
where the deterministic weights $a_{n,i}$s satisfy
$$\max_{1\leq i\leq n}\frac{a_{n,i}}{\sqrt n}\rightarrow 0\quad \text{and} \quad \frac1n \sum_{i=1}^{n-1}a_{n,i}^2\rightarrow \int_0^1\{\log(1/t)\}^2dt=2.$$
Moreover, $\E(\boldsymbol H_i\mid\mathcal F_{i-1})=\boldsymbol0_K$ and, since the off-diagonal conditional covariances vanish (only one treatment can be assigned to each subject), it follows that
$$\E\left(\boldsymbol H_i\boldsymbol H_i^\top \mid\mathcal F_{i-1}\right)=\operatorname{diag}\left(\frac{\Lambda_k(\widehat{\boldsymbol{\Psi}}_{i-1})\E[\varphi_k(Y_k)^2]}{\Lambda_k(\boldsymbol\Psi)^2}\right)_{k=1,\ldots,K}\rightarrow\boldsymbol\Sigma\qquad\text{a.s.}$$
so that
$$\frac1n \sum_{i=1}^{n-1}a_{n,i}^2 \boldsymbol J\E\left(\boldsymbol H_i\boldsymbol H_i^\top\mid\mathcal F_{i-1}\right)\boldsymbol J^\top
\rightarrow 2\boldsymbol J \boldsymbol\Sigma \boldsymbol J^\top\qquad\text{a.s.}$$
Furthermore, $\E\left(\boldsymbol e_i\boldsymbol e_i^\top\mid\mathcal F_{i-1}\right)=\operatorname{diag}\left\{\boldsymbol\Lambda(\widehat{\boldsymbol{\Psi}}_{i-1})\right\}
-\boldsymbol\Lambda(\widehat{\boldsymbol{\Psi}}_{i-1})\boldsymbol\Lambda(\widehat{\boldsymbol{\Psi}}_{i-1})^\top$ and
$$\frac1n\sum_{i=1}^n \E\left(\boldsymbol e_i\boldsymbol e_i^\top \mid\mathcal F_{i-1}\right)\rightarrow
\operatorname{diag}\left\{\boldsymbol\Lambda(\boldsymbol\Psi)\right\}-\boldsymbol\Lambda(\boldsymbol\Psi)\boldsymbol\Lambda(\boldsymbol\Psi)^\top \qquad\text{a.s.}$$
Finally, for every pair $k,j\in\{1,\ldots,K\}$, $\E\left(e_{i,k}H_{i,j}\mid\mathcal F_{i-1}\right)=0$, because the current treatment assignment is conditionally independent
of the current potential outcomes and $\E\{\varphi_j(Y_j)\}=0$. Therefore, the asymptotic distribution in (\ref{CLTpi}) follows from (\ref{decomp2}) via the martingale central limit theorem (the Lindeberg condition holds from \ref{ass:CLT2} and since $\max_i a_{n,i}/\sqrt n\to0$).

\subsection{Proof of Theorem \ref{th:optimal_scores}}\label{proofTHM4}
Under $A$-optimality, the optimal target minimizes
$$\min_{\boldsymbol{\rho}\in\mathcal{S}_K}\operatorname{tr}\{\boldsymbol{V}(\boldsymbol{\rho})\}=\min_{\boldsymbol{\rho}\in\mathcal{S}_K}\sum_{k=1}^K\frac{\nu_k^2}{\rho_k}.$$
By the Cauchy-Schwarz inequality,
$$\left(\sum_{k=1}^K\nu_k\right)^2\leq \left(\sum_{k=1}^K\frac{\nu_k^2}{\rho_k}\right)\left(\sum_{k=1}^K\rho_k\right)=\sum_{k=1}^K\frac{\nu_k^2}{\rho_k},$$
with equality if and only if $\rho_k\propto\nu_k$, yielding the Neyman-type allocation $\rho_k^{A}= \nu_k/(\sum_{j=1}^K\nu_j)$, for $k=1,\ldots,K$. $D$-optimality instead minimizes
$$\min_{\boldsymbol{\rho}\in\mathcal{S}_K}\det\{\boldsymbol{V}(\boldsymbol{\rho})\} =\left(\prod_{k=1}^K\nu_k^2\right)\min_{\boldsymbol{\rho}\in\mathcal{S}_K}\prod_{k=1}^K\rho_k^{-1},$$
leading to the balanced target $\rho_k^{D}=K^{-1}$, for $k=1,\ldots,K$. Finally, $E$-optimality minimizes
$$\min_{\boldsymbol{\rho}\in\mathcal{S}_K}\lambda_{\max}\{\mathbf{V}(\boldsymbol{\rho})\}=\min_{\boldsymbol{\rho}\in\mathcal{S}_K}\max_{k=1,\ldots,K}\frac{\nu_k^2}{\rho_k}$$
If $M=\max_k\nu_k^2/\rho_k$, then $\nu_k^2\leq M\rho_k$ for every $k$ and hence $\sum_{k=1}^K\nu_k^2\leq M$, where equality is attained by taking
$\rho_k^{E}=\nu_k^2/\sum_{j=1}^K\nu_j^2$.

\subsection{Proof of Theorem \ref{thmERADE}}\label{proofTHM3}
From (\ref{NPERADE}), $\Pr(\delta_{n+1,k}=1\mid\mathcal F_n)\geq \gamma\widehat\rho_{n,k}$ for $k=1,\ldots,K$. Thus, \ref{assERADE1} and the conditional Borel-Cantelli lemma imply that $N_{n,k}\to\infty$ a.s. for every treatment $k$. By the independence-after-adaptive-allocation property introduced in Section \ref{Sec2}, the Glivenko-Cantelli theorem then yields $\|\widehat F_{n,k}-F_k\|_\infty\to0 $ a.s. for every $k$. Hence, by \ref{assERADE1}, $\widehat{\boldsymbol{\rho}}_n\to\boldsymbol{\rho}$ a.s. and therefore the standard multi-arm ERADE tracking argument yields $\boldsymbol{\pi}_n\to\boldsymbol{\rho}$ a.s. Thus, the treatment-specific empirical-process limit implied by the independence-after-adaptive-allocation property gives $\sqrt{N_{n,k}}(\widehat F_{n,k}-F_k)\Rightarrow \mathbb{G}_k\circ F_k$, for every $k=1,\ldots,K$. Since $\pi_{n,k}=N_{n,k}/n\to\rho_k>0$ a.s. and
$$\sqrt n(\widehat F_{n,k}-F_k)=\pi_{n,k}^{-1/2}\sqrt{N_{n,k}}(\widehat F_{n,k}-F_k),$$
Slutsky's theorem yields \eqref{ERADE-FCLT}, while \eqref{ERADE-score} follows from the functional delta method.

If, in addition, condition \ref{assERADE2} holds, its asymptotic linear representation is of the same type as the Bahadur expansion used in the multi-arm ERADE analysis of \cite{Hu2025}. The remainder in (\ref{rho-linear}) satisfies
$\max_{1\leq \ell\leq n}\ell\|\boldsymbol r_\ell^\rho-\boldsymbol r_n^\rho\| \leq  2\max_{1\leq m\leq n}m\|\boldsymbol r_m^\rho\|=o_{\mathrm{a.s.}}(\sqrt n)$.
Thus, the remainder is negligible uniformly over the random indices appearing in their tracking argument. Together with the already established convergence $\boldsymbol\pi_n\to\boldsymbol\rho$ a.s., this provides the controls needed to repeat the root-$n$ tracking argument in \cite{Hu2025}, yielding $\boldsymbol N_n-n\widehat{\boldsymbol{\rho}}_n=o_p(n^{1/2})$, which proves \eqref{ERADE-tracking}.

As regards the estimated target, from \ref{assERADE2},
$$\sqrt n (\widehat{\boldsymbol{\rho}}_n-\boldsymbol{\rho}) =\sum_{k=1}^K \pi_{n,k}^{-1/2} \frac{1}{\sqrt{N_{n,k}}} \sum_{i=1}^n\delta_{ik}\boldsymbol h_{\rho,k}(Y_{ik}) + o_p(1).$$
By the independence-after-adaptive-allocation argument,
$$ \frac{1}{\sqrt{N_{n,k}}}\sum_{i=1}^n \delta_{ik}\boldsymbol h_{\rho,k}(Y_{ik})\Rightarrow N\left( \boldsymbol 0_K,\E\left[\boldsymbol{h}_{\rho,k}(Y_k)
\boldsymbol{h}_{\rho,k}(Y_k)^\top\right]\right),$$
jointly across treatments, with independent limiting components. Since $\pi_{n,k}\to\rho_k$ a.s., Slutsky's theorem yields the asymptotic normality of $\sqrt n(\widehat{\boldsymbol\rho}_n-\boldsymbol\rho)$; combining this result with \eqref{ERADE-tracking} gives \eqref{ERADE-target}.

Finally, under Condition \ref{assERADE3}, $\boldsymbol{h}_{\rho,k}$ is the canonical-gradient component associated with perturbations of $F_k$. Since the treatment-assignment probabilities are known predictable functions of the observed history, the randomization mechanism does not contribute to the likelihood score. Since $\boldsymbol\pi_n\to\boldsymbol\rho$ a.s. and $n^{-1}\sum_{i=1}^n \{\delta_{ik}-\Pr(\delta_{ik}=1\mid\mathcal F_{i-1})\} \rightarrow 0 $ a.s. for every $k$, the martingale strong law yields $n^{-1}\sum_{i=1}^n \Pr(\delta_{ik}=1\mid\mathcal F_{i-1}) \rightarrow\rho_k$ a.s. Consequently, for any collections of scores $\boldsymbol s=(s_1,\ldots,s_K)$ and $\boldsymbol t=(t_1,\ldots,t_K)$, the inner product in the limiting experiment is
$$\langle\boldsymbol s,\boldsymbol t\rangle_{\boldsymbol\rho} =\sum_{k=1}^K\rho_k \E\{s_k(Y_k)t_k(Y_k)\}.$$
By \ref{assERADE3}, the pathwise derivative of the target in the direction $\boldsymbol s$ is $\sum_{k=1}^K\E\{\boldsymbol h_{\rho,k}(Y_k)s_k(Y_k)\}$ and its Riesz representer has treatment-specific components $\boldsymbol h_{\rho,k}/\rho_k$; the corresponding efficient influence-function contribution of the $i$th observation is
$$\boldsymbol\varphi_{\rho,i}=\sum_{k=1}^K\frac{\delta_{ik}}{\rho_k}\boldsymbol h_{\rho,k}(Y_{ik}).$$
Moreover,
$$\frac{1}{n}\sum_{i=1}^n\E\left(\boldsymbol\varphi_{\rho,i}\boldsymbol\varphi_{\rho,i}^{\top}\mid\mathcal F_{i-1}\right)\rightarrow \sum_{k=1}^K \rho_k^{-1}
\E\left\{\boldsymbol h_{\rho,k}(Y_k)\boldsymbol h_{\rho,k}(Y_k)^{\top}\right\}=\boldsymbol\Omega \qquad\text{a.s.}$$
which is the semiparametric efficiency bound for estimating $\boldsymbol{\rho}$. From \eqref{ERADE-tracking}, $\boldsymbol{\pi}_n$ and $\widehat{\boldsymbol{\rho}}_n$ are asymptotically equivalent at the $n^{-1/2}$ scale, thus the NP-ERADE allocation proportions attain the same bound.

\bibliographystyle{spr-ims-number}
\bibliography{Biblio6}
\end{document}